\documentclass[11pt,a4paper]{article}
\pdfoutput=1

\usepackage{jheppub}

\usepackage{amsmath}
\usepackage{bbm}
\usepackage{float}
\usepackage{graphicx}	
\usepackage{hyperref}
\usepackage{mathtools}
\usepackage{cancel}

\usepackage[normalem]{ulem}

\usepackage{multirow}
\usepackage{rotating}
\usepackage{subcaption}

\def\lsim{\raise0.3ex\hbox{$\;<$\kern-0.75em\raise-1.1ex
\hbox{$\sim\;$}}}
\def\gsim{\raise0.3ex\hbox{$\;>$\kern-0.75em\raise-1.1ex
\hbox{$\sim\;$}}}

\def\thetitle{ 
Symmetry Finder applied to the 1-3 mass eigenstate exchange symmetry \\
 \vspace{- 6mm}
}
\title{\thetitle}
\hypersetup{pdftitle={\thetitle}}
\author{Hisakazu Minakata}
\affiliation{
Center for Neutrino Physics, Department of Physics, Virginia Tech, Blacksburg, Virginia 24061, USA \\
}

\emailAdd{hisakazu.minakata@gmail.com}
\date{\today}

\abstract{
In a previous paper, {\em Symmetry Finder} (SF) method is proposed to find the reparametrization symmetry of the state-exchange type in neutrino oscillation in matter. It has been applied successfully to the 1-2 state exchange symmetry in the DMP perturbation theory, yielding the eight symmetries. In this paper, we apply the SF method to the atmospheric-resonance perturbation theory to uncover the 1-3 state relabeling symmetries. The pure 1-3 state symmetry takes the unique position that it is practically impossible to formulate in vacuum under the conventional choice of the flavor mixing matrix. In contrast, our SF method produces the sixteen 1-3 state exchange symmetries in matter. The relationship between the symmetries in the original (vacuum plus matter) Hamiltonian and the ones in the diagonalized system is discussed. 

}

\begin{document} 

\maketitle

\section{Introduction}
\label{sec:introduction} 

Symmetry consideration plays an important role in understanding the system in quantum mechanics and quantum field theory~\cite{Coleman:1985rnk}. It should be true in neutrino oscillation, which plays a crucial role in the measurement of the flavor mixing angles, the CP phase, neutrino masses and the mass patterns~\cite{Zyla:2020zbs}. These informations have been and will be the source of stimulation for physics of neutrino masses and the flavor mixing~\cite{Mohapatra:2006gs,Altarelli:2010gt}. 
Naturally, various symmetries and mapping properties are discussed in many different contexts, possibly including the parameter space alternative to the customary assumed ones, or interactions beyond the neutrino-mass embedded Standard Model ($\nu$SM)~\cite{Fogli:1996nn,deGouvea:2000pqg,Fogli:1996pv,Fogli:2001wi,Minakata:2001qm,Minakata:2010zn,deGouvea:2008nm,Coloma:2016gei,Denton:2016wmg,Zhou:2016luk,Martinez-Soler:2019nhb,Minakata:2020oxb}.\footnote{
It is likely that we miss many other relevant references. } 

In a previous paper~\cite{Minakata:2021dqh}, we have proposed a systematic method for finding symmetry in neutrino oscillation probability in matter, which is dubbed as {\em Symmetry Finder} (SF). By symmetry we mean invariance under a state-relabeling and the associated redefinitions of the mixing angles, or inclusively under the reparametrization. The SF method starts from the observation that the two different expressions of the flavor state in terms of the energy eigenstate imply a symmetry~\cite{Parke:2018shx}. If one of the expressions contains the 1-2 state exchange, for example, the appearing symmetry is of the 1-2 state exchange type. Often the rephasing of either one of the flavor or the energy eigenstates, or the both are involved. Because of its efficient function of hunting  symmetries, the flavor - mass eigenstates relation is termed as the SF equation. 

Despite its simple structure in vacuum, the application of the SF method in matter environments requires a new formalism which is set up in ref.~\cite{Minakata:2021dqh}. The resultant machinery for uncovering the state-relabeling, or reparametrization symmetry in neutrino oscillation in matter has been applied to the Denton {\it et al.} (DMP) perturbation theory~\cite{Denton:2016wmg}. The SF method has proved to be powerful as it produced the eight 1-2 state exchange symmetries~\cite{Minakata:2021dqh}, all of which are new except for the unique exception uncovered in ref.~\cite{Denton:2016wmg}. Since only a few of the similar symmetries were known~\cite{Martinez-Soler:2019nhb} before ref.~\cite{Minakata:2021dqh}, it is fair to say that the SF method is powerful and successful, to the opinion of the present author. 

In this paper we discuss the 1-3 state exchange reparametrization symmetry in matter by using the SF method. This is the first systematic treatment of the 1-3 exchange symmetry to our knowledge. The 1-3 relabeling symmetry takes a very different position from the 1-2 exchange symmetry. As we will briefly mention in section~\ref{sec:overview}, there is an ongoing discussion on how to understand the relationship between the vacuum symmetry, invariance under the vacuum variable transformations, and the symmetry written by the dynamical, or the matter-dressed variables. What is unique in the 1-3 exchange symmetry is that it is practically impossible to write down the pure 1-3 state relabeling symmetry in vacuum, the symmetry in which only the 1 and 3 states are involved. On the other hand, our treatment using the SF method produces the sixteen 1-3 state relabeling symmetry in matter, as we will see in sections~\ref{sec:13-symmetry} and \ref{sec:13-symmetry-result}. The key requisite is the use of the ``correct'' perturbative framework, the atmospheric-resonance perturbation theory~\cite{Minakata:2015gra}, in this case. We hope that the new series of the 1-3 exchange symmetry contributes a better understanding of the state relabeling symmetry in neutrino oscillation in matter. 

What is the significance of the state-relabeling, or reparametrization symmetry in neutrino oscillation? The question arises probably because the state-relabeling is an operation done inside the  theoretical treatment, and it does not appear to carry an obvious physical meaning. While this itself is true, we will learn through investigation of reparametrization symmetry in matter that our theoretical understanding of the three-flavor neutrino oscillations is far below the matured level. For example, we do not know how big is the reparametrization symmetry in the system. Therefore, our symmetry discussion serves for diagnostic of neutrino theory. We will bring the readers to a door open for further discussions on understanding the nature of the state-relabeling symmetry in section~\ref{sec:overview}. On the other side of the question, about the practical utility, these symmetries serve as a useful tool to make a consistency check of the derived expressions of the oscillation probabilities. 

\section{The three neutrino evolution in matter in the $\nu$SM and its diagonalization}
\label{sec:3nu-SM} 

Though standard by based on $\nu$SM, we first define our system, evolution of the three-flavor neutrinos defined by the Hamiltonian $H$ in the flavor basis. In this paper we often discuss the two ways of expressing the Hamiltonian $H$, the originally defined form which will be denoted as $H_{ \text{LHS} }$, and its diagonalized form $H_{ \text{RHS} }$.\footnote{
Throughout this paper, ``LHS'' and ``RHS'' are shorthand of the terms ``left-hand side'' and ``right-hand side'', respectively. 
}
Of course, they must be equal to each other, $H_{ \text{LHS} } = H_{ \text{RHS} }$. They are expressed after multiplication of $2E$ with $E$ being neutrino energy, respectively, as 
\begin{eqnarray} 
&& 
2E H_{ \text{LHS} } 
= 
U_{23} (\theta_{23}) U_{13} (\theta_{13}, \delta) U_{12} (\theta_{12}) 
\left[
\begin{array}{ccc}
m^2_{1} & 0 & 0 \\
0 & m^2_{2} & 0 \\
0 & 0 & m^2_{3}
\end{array}
\right] 
U_{12}^{\dagger} (\theta_{12}) U_{13}^{\dagger} (\theta_{13}, \delta) U_{23}^{\dagger} (\theta_{23})
\nonumber \\
&&
\hspace{16mm} 
+ \left[
\begin{array}{ccc}
a(x) & 0 & 0 \\
0 & 0 & 0 \\
0 & 0 & 0
\end{array}
\right], 
\label{H-LHS}
\end{eqnarray}
\begin{eqnarray}
&&
\hspace{-4mm}
2E H_{ \text{RHS} } 
=
U_{23} (\widetilde{\theta_{23}}) 
U_{13} (\widetilde{\theta_{13}}, \widetilde{\delta})
U_{12} (\widetilde{\theta_{12}}) 
\left[
\begin{array}{ccc}
\lambda_{1} & 0 & 0 \\
0 & \lambda_{2} & 0 \\
0 & 0 & \lambda_{3} \\
\end{array}
\right] 
U_{12}^{\dagger} (\widetilde{\theta_{12}}) 
U_{13}^{\dagger} (\widetilde{\theta_{13}}, \widetilde{\delta}) 
U_{23}^{\dagger} (\widetilde{\theta_{23}}). 
\label{H-RHS}
\end{eqnarray} 

In eq.~\eqref{H-LHS}, $U \equiv U_{12}^{\dagger} (\theta_{12}) U_{13}^{\dagger} (\theta_{13}, \delta) U_{23}^{\dagger} (\theta_{23})$ denotes the standard $3 \times 3$ lepton flavor mixing matrix~\cite{Maki:1962mu} in the Particle Data Group (PDG) convention~\cite{Zyla:2020zbs}, which relates the flavor neutrino states to the vacuum mass eigenstates as $\nu_{\alpha} = U_{\alpha i} \nu_{i}$, where $\alpha$ runs over $e, \mu, \tau$, and the mass eigenstate index $i$ runs over $1,2,$ and $3$. Our notations for the mixing angles and the CP phase (i.e., lepton Kobayashi-Maskawa phase~\cite{Kobayashi:1973fv}) are the standard ones.
The functions $a(x)$ in eq.~\eqref{H-LHS} denote the Wolfenstein matter potential \cite{Wolfenstein:1977ue} due to the charged current (CC) reactions 
\begin{eqnarray} 
a (x) &=&  
2 \sqrt{2} G_F N_e E \approx 1.52 \times 10^{-4} \left( \frac{Y_e \rho (x) }{\rm g\,cm^{-3}} \right) \left( \frac{E}{\rm GeV} \right) {\rm eV}^2.  
\label{matt-potential}
\end{eqnarray}
Here, $G_F$ is the Fermi constant, $N_e$ is the electron number density in matter. $\rho (x)$ and $Y_e$ denote, respectively, the matter density and number of electron per nucleon in matter. 

In $H_{ \text{RHS} }$ in eq.~\eqref{H-RHS}, $\lambda_{i}$ ($i=1,2,3$) are the eigenvalues in matter, and $\widetilde{\theta_{12}}$, $\widetilde{\delta}$, and etc. denote the mixing angles and CP phase in matter. The expressions of $\lambda_{i}$ are obtained in ref.~\cite{Barger:1980tf}, and the matter mixing angles and phases by Zaglauer and Schwarzer (ZS)~\cite{Zaglauer:1988gz}, both under the uniform matter density approximation. 
For notational convenience we denote the first and the second terms of $H_{ \text{LHS} }$ as $H_{ \text{vac} }$ and $H_{ \text{matt} }$, respectively. 

\subsection{Symmetry in $H_{ \text{LHS} }$ vs. Symmetry in $H_{ \text{RHS} }$}
\label{sec:LHS-RHS} 

In ref.~\cite{Minakata:2021dqh} the vacuum symmetry of 1-2 mass-eigenstates exchange type is reviewed. It can be shown that the SF equation in vacuum leads to the following two symmetries~\cite{Parke:2018shx}
\begin{eqnarray} 
&& 
\text{Symmetry IA-vacuum:}
\hspace{6mm}
m^2_{1} \leftrightarrow m^2_{2}, 
\hspace{6mm} 
\cos \theta_{12} \rightarrow \mp \sin \theta_{12},
\hspace{6mm}
\sin \theta_{12} \rightarrow \pm \cos \theta_{12},
\nonumber \\
&&
\text{Symmetry IB-vacuum:}
\hspace{6mm}
m^2_{1} \leftrightarrow m^2_{2}, 
\hspace{8mm}
\cos \theta_{12} \leftrightarrow \sin \theta_{12}, 
\hspace{8mm}
\delta \rightarrow \delta \pm \pi.
\label{Symmetry-IA-IB-vacuum}
\end{eqnarray} 
In a nutshell, the two different expressions of neutrino flavor state by the mass eigenstate,\footnote{
Here we use the SOL convention of the flavor mixing matrix $U$, in which $e^{ \pm i \delta}$ is attached to $s_{12}$. For its relation to the PDG and the other conventions, see ref.~\cite{Martinez-Soler:2018lcy}.
}
\begin{eqnarray} 
&& 
\left[
\begin{array}{c}
\nu_{e} \\
\nu_{\mu} \\
\nu_{\tau} \\
\end{array}
\right] 
=
U_{23} (\theta_{23}) U_{13} (\theta_{13}) 
U_{12} (\theta_{12}, \delta) 
\left[
\begin{array}{c}
\nu_{1} \\
\nu_{2} \\
\nu_{3} \\
\end{array}
\right] 
=
U_{23} (\theta_{23}) U_{13} (\theta_{13}) 
U_{12} \left( \theta_{12} + \frac{\pi}{2}, \delta \right) 
\left[
\begin{array}{c}
- e^{ i \delta } \nu_{2}  \\
e^{ - i \delta } \nu_{1} \\
\nu_{3} \\
\end{array}
\right], 
\nonumber \\
\label{flavor-mass-vacuum}
\end{eqnarray}
implies the symmetry under the transformations we referred as Symmetry IA-vacuum (upper sign) in eq.~\eqref{Symmetry-IA-IB-vacuum}. The other flavor - mass eigenstate relation can be written down and leads to Symmetry IB-vacuum which contains the $\delta$ transformation, see refs.~\cite{Parke:2018shx,Minakata:2021dqh}. 
Notice that they can be regarded as the symmetries of the total Hamiltonian eq.~\eqref{H-LHS}, because the matter potential term is obviously invariant under the transformations in eq.~\eqref{Symmetry-IA-IB-vacuum}. 

Since the Hamiltonian of the ZS system~\eqref{H-RHS} has the same form as $H_{ \text{vac} }$ with replacements $m^2_{j} \rightarrow \lambda_{j}$, $\theta_{ij} \rightarrow \widetilde{\theta_{ij}}$ and $\delta \rightarrow \widetilde{\delta}$, it has the matter version of the above vacuum symmetries, termed as IA-ZS and IB-ZS in ref.~\cite{Minakata:2021dqh}:
\begin{eqnarray} 
&& 
\text{Symmetry IA-ZS:}
\hspace{6mm}
\lambda_{1} \leftrightarrow \lambda_{2}, 
\hspace{6mm} 
\cos \widetilde{\theta_{12}} \rightarrow \mp \sin \widetilde{\theta_{12}},
\hspace{6mm}
\sin \widetilde{\theta_{12}} \rightarrow \pm \cos \widetilde{\theta_{12}}, 
\nonumber \\
&&
\text{Symmetry IB-ZS:}
\hspace{6mm}
\lambda_{1} \leftrightarrow \lambda_{2}, 
\hspace{6mm} 
\cos \widetilde{\theta_{12}} \leftrightarrow \sin \widetilde{\theta_{12}},
\hspace{10mm}
\tilde{\delta} \rightarrow \tilde{\delta} \pm \pi.
\label{Symmetry-IA-IB-ZS} 
\end{eqnarray} 
Notice that these are the symmetries whose transformations consist only of the matter variables. None of the vacuum parameters in $H_{ \text{RHS} }$ in eq.~\eqref{H-RHS} transforms. 

What is the nature of the symmetries Symmetry IA-ZS and IB-ZS in matter? 
What is the interpretation of symmetries in the DMP system~\cite{Denton:2016wmg}, as well as the ones found in ref.~\cite{Martinez-Soler:2019nhb}? We have made several remarks about this question. The first one is that they are the ``dynamical symmetry''~\cite{Martinez-Soler:2019nhb,Martinez-Soler:2018lcy}.\footnote{
A dynamical symmetry is the symmetry that has no obvious trace in the Hamiltonian of the system, but the one which indeed arises after the system is solved. The symmetry often involves the variables that are used to diagonalize the Hamiltonian. 
}
The second characterization we have coined is that these symmetries arise due to the rephasing invariance of the $S$ matrix~\cite{Minakata:2020oxb}. We believe that both of the interpretations are still valid, illuminating the alternative aspects. 

A note for the nomenclature: In this paper we denote a symmetry of the Hamiltonian $H_{ \text{LHS} }$ in eq.~\eqref{H-LHS} as ``vacuum symmetry'', and a symmetry of the Hamiltonian $H_{ \text{RHS} }$ in eq.~\eqref{H-RHS} as ``matter symmetry''. The vacuum symmetry transformations are written by the masses and the mixing parameters in vacuum, whereas the matter symmetry transformations are expressed by using the Hamiltonian-diagonalizing or matter-dressed variables. The Hamiltonian $H_{ \text{RHS} }$ need not to be the completely diagonalized form as in eq.~\eqref{H-RHS}. It can be the form of a sum of nearly diagonalized term plus corrections. See section~\ref{sec:hamiltonian-view}.


\section{Vacuum symmetry vs. matter symmetry}
\label{sec:overview}

It may be illuminative to briefly discuss the relationship between the vacuum symmetry and matter symmetry to understand the nature of state relabeling symmetries in neutrino oscillations in matter. 

\subsection{Vacuum symmetry approach}
\label{sec:VSA}

We can start from the following argument: Suppose that one finds a symmetry X of $H_{ \text{vac} }$, the first term in $H_{ \text{LHS} }$ in eq.~\eqref{H-LHS}, whose transformations consist of the vacuum parameters. Then, the symmetry X must be the symmetry of $H_{ \text{LHS} }$ because the matter term $H_{ \text{matt} }$ does not transform by X~\cite{Fogli:1996nn,Zhou:2016luk}. Then, $H_{ \text{RHS} }$ must also be invariant under X because they are equal, $H_{ \text{RHS} } = H_{ \text{LHS} }$. If this argument is valid, one can find symmetries of $H_{ \text{RHS} }$ by using $H_{ \text{LHS} }$ only. 

\subsection{Two-flavor model and its vacuum symmetry} 
\label{sec:2-flavor}

We analyze a two-flavor model of neutrino oscillations to know how symmetry of $H_{ \text{LHS} }$ can reveal symmetry of $H_{ \text{RHS} }$.\footnote{
The author thanks the referee of EPJC for the comments which led him to the treatment of the two-flavor model in this section. } 
In this model, whose Hamiltonian is the two-flavor version of eq.~\eqref{H-LHS}, the flavor and the mass eigenstates are related by $\nu_{\alpha} = U_{\alpha j} (\theta) \nu_{j}$ ($\alpha = e, \mu$, $j=1,2$) where $U (\theta)$ denotes the two-dimensional rotation matrix with the vacuum mixing angle $\theta$. We assume that $m^2_{2} > m^2_{1}$. Then, $2E$ times the flavor-basis Hamiltonian reads 
\begin{eqnarray} 
&& 
\hspace{-4mm}
2E H_{ \text{LHS} } = 
U (\theta)
\left[
\begin{array}{cc}
m^2_{1} & 0 \\
0 & m^2_{2} \\
\end{array}
\right] 
U (\theta)^{\dagger}
+ 
\left[
\begin{array}{cc}
a & 0 \\
0 & 0 \\
\end{array}
\right] 
= \left[
\begin{array}{cc}
\cos^2 \theta m^2_{1} + \sin^2 \theta m^2_{2} + a & \cos \theta \sin \theta \Delta m^2 \\
\cos \theta \sin \theta \Delta m^2 & \sin^2 \theta m^2_{1} + \cos^2 \theta m^2_{2} \\
\end{array}
\right], 
\nonumber \\
\label{2EH-LHS}
\end{eqnarray}
where $\Delta m^2 \equiv m^2_{2} - m^2_{1}$, and $a$ denotes Wolfenstein's potential for uniform density matter. We observe that $H_{ \text{LHS} }$ is invariant under the transformations of Symmetry IA-vacuum, $\theta_{12} \rightarrow \theta$ in eq.~\eqref{Symmetry-IA-IB-vacuum}. 

We formulate the small-$\theta$ perturbation theory in a way keeping the manifest invariance under Symmetry IA-vacuum. For this purpose we make an overall phase redefinition of the neutrino state such that the unit matrix 
$( \cos^2 \theta m^2_{1} + \sin^2 \theta m^2_{2} ) \bf{1}$ is subtracted from the Hamiltonian~\eqref{2EH-LHS}~\cite{Zhou:2016luk}. Then, we decompose the Hamiltonian into the unperturbed and perturbed parts, $2E H_{ \text{LHS} } = 2E H_{0} + 2E H_{1}$, where 
\begin{eqnarray} 
&& 
2E H_{0}
= 
\left[
\begin{array}{cc}
a & 0 \\
0 & \cos 2\theta \Delta m^2 \\
\end{array}
\right], 
\hspace{10mm}
2E H_{1}
= 
\left[
\begin{array}{cc}
0 & \frac{1}{2} \sin 2\theta \Delta m^2 \\
\frac{1}{2} \sin 2\theta \Delta m^2 & 0 \\
\end{array}
\right]. 
\label{H0-H1}
\end{eqnarray}
Notice that the both $2E H_{0}$ and $2E H_{1}$ are separately invariant under Symmetry IA. In this way the small-$\theta$ perturbation theory can be formulated in such a way that invariance under Symmetry IA-vacuum transformations is manifest in each order in perturbation theory~\cite{Zhou:2016luk}. 

\subsection{Matter symmetry vs. perturbative vacuum symmetry} 
\label{sec:matter-vs-vacuum}

We now show that the symmetry of $H_{ \text{RHS} }$ is {\em not} identical with Symmetry IA-vacuum, even though it is respected in each order in perturbation theory. The Hamiltonian~\eqref{2EH-LHS} can be diagonalized by the rotation with the matter angle $\widetilde{\theta}$, which yields the eigenvalues of $2E H_{ \text{LHS} }$ and $\widetilde{\theta}$ as 
\begin{eqnarray} 
&& 
\lambda_{2, 1}
= \frac{1}{2} 
\biggl\{ 
( a + m^2_{1} + m^2_{2} ) 
\pm 
\sqrt{ a^2 - 2 a \cos 2\theta \Delta m^2  + ( \Delta m^2 )^2 } 
\biggr\}, 
\nonumber \\
&& 
\cos 2 \widetilde{\theta}
= \frac{ \cos 2\theta \Delta m^2 - a  } 
{ \lambda_{2} - \lambda_{1} }, 
\hspace{10mm}
\sin 2 \widetilde{\theta}
= 
\frac{ \sin 2\theta \Delta m^2  } 
{ \lambda_{2} - \lambda_{1} }. 
\label{2-nu-eigenvalue-angle}
\end{eqnarray}
The $\pm$ sign in the eigenvalues are taken such that $\lambda_{2} > \lambda_{1}$. Then, there is no sign ambiguity in $\cos 2 \widetilde{\theta}$ and $\sin 2 \widetilde{\theta}$, because it must be that $\widetilde{\theta} \rightarrow \theta$ as $a \rightarrow 0$ and $\widetilde{\theta} \rightarrow \frac{\pi}{2}$ as $a \rightarrow + \infty$, the usual MSW mechanism~\cite{Mikheyev:1985zog,Wolfenstein:1977ue}.
The diagonalized Hamiltonian takes the form $H_{ \text{RHS} } = U (\widetilde{\theta}) \text{diag} (\lambda_{1}, \lambda_{2}) U (\widetilde{\theta})^{\dagger}$, and the oscillation probability is given by 
\begin{eqnarray} 
&& 
P (\nu_{\mu} \rightarrow \nu_{e}) 
= 
\sin^2 2\widetilde{\theta} 
\sin^2 \frac{ (\lambda_{2} - \lambda_{1}) L }{4E}, 
\label{P-mue}
\end{eqnarray}
where $L$ is the baseline. By using the matter variable expressions of the eigenvalues 
\begin{eqnarray} 
&&
\lambda_{1} 
= 
\cos^2 \widetilde{\theta} a + \frac{1}{2} ( m^2_{2} + m^2_{1} ) 
- \frac{1}{2} \cos 2 ( \widetilde{\theta} - \theta ) \Delta m^2, 
\nonumber \\
&& 
\lambda_{2} 
= 
\sin^2 \widetilde{\theta} a + \frac{1}{2} ( m^2_{2} + m^2_{1} ) 
+ \frac{1}{2} \cos 2 ( \widetilde{\theta} - \theta ) \Delta m^2, 
\label{2-nu-eigenvalue-matter}
\end{eqnarray}
the transformations of $\widetilde{\theta}$ in eq.~\eqref{2-nu-eigenvalue-angle} are consistent with $\lambda_{1} \leftrightarrow \lambda_{2}$. In the both expressions eqs.~\eqref{2-nu-eigenvalue-angle}, or \eqref{2-nu-eigenvalue-matter}, it follows that $\lambda_{j} \rightarrow m^2_{j}$ ($j=1,2$) in the vacuum limit $a \rightarrow 0$. 

Therefore, the symmetry of $H_{ \text{RHS} }$ is the two-flavor version of the matter symmetry IA-ZS, see eq.~\eqref{Symmetry-IA-IB-ZS}. A natural question would be why it is different from Symmetry IA-vacuum, which is concluded in the perturbative approach in section~\ref{sec:2-flavor}. 

To understand the point we must notice first that both the eigenvalues $\lambda_{j}$ and the matter angle $\widetilde{\theta}$ are invariant under the transformations of 
Symmetry IA-vacuum, the symmetry of $H_{ \text{LHS} }$~\cite{Denton:2021vtf}. The diagonalized Hamiltonian $H_{ \text{RHS} }$ describes the dynamics of the system, and it must be independent of how (in which way) the vacuum Hamiltonian, and hence $H_{ \text{LHS} }$, is parametrized. Therefore, the vacuum relabeling symmetry cannot affect the physical system, and hence the symmetry of $H_{ \text{RHS} }$.\footnote{
For a different view that the symmetry of $H_{ \text{RHS} }$ is the product, IA-vacuum $\times$ IA-ZS, see ref.~\cite{Denton:2021vtf}. }
It is what our two-flavor toy model reveals. Hence, it appears that the vacuum symmetry approach fails to identify the system's matter symmetry. 

\subsection{All-order summation of perturbative series} 
\label{sec:all-order}

However, the picture changes when the perturbative series is summed to all orders.\footnote{
Assuming convergence in a small radius one can show that the series can be exponentiated. The result may be analytically continued to a larger domain of the expansion parameter. 
}
Obviously it reproduces the system with $\lambda_{2,1}$ (apart from the constant shift of $\cos^2 \theta m^2_{1} + \sin^2 \theta m^2_{2}$) and $\widetilde{\theta}$ in \eqref{2-nu-eigenvalue-angle}. The symmetry of the system is clearly the two-flavor version of the matter symmetry IA-ZS. Thus, we have arrived at a rather complicated, or profound, picture of the relationship between the symmetries of $H_{ \text{LHS} }$ and $H_{ \text{RHS} }$: 
\begin{itemize}

\item 
In any finite order in the small-$\theta$ perturbation theory using the bases in eq.~\eqref{H0-H1}, the symmetry of the system predicted by $H_{ \text{LHS} }$ is Symmetry IA-vacuum. 

\item 
When all orders are summed, the symmetry of the system becomes Symmetry IA-ZS. It appears that Symmetry IA-vacuum fuses into IA-ZS, the non-perturbative matter symmetry of $H_{ \text{RHS} }$. 

\end{itemize}
The emerged feature suggests that Symmetry IA-ZS is not completely independent of Symmetry IA-vacuum, reflecting the fact that $H_{ \text{LHS} }$ and $H_{ \text{RHS} }$ define the same theory. Even though the readers might feel the above picture contrived one, it seems to be the reality which is extracted from an explicit treatment of the two-flavor model.

\section{General constraints on state exchange symmetry} 
\label{sec:G-constraints} 

In section~\ref{sec:overview}, we have used the two-flavor model to investigate the relationship between the 1-2 state exchange symmetries in $H_{ \text{LHS} }$ and $H_{ \text{RHS} }$. To extend the similar consideration to the three-flavor system, we consider the possibility that the general relations which reflect the equality $H_{ \text{LHS} } = H_{ \text{RHS} }$ might be useful. 
In fact, there exist the identities \'a la Naumov~\cite{Naumov:1991ju} and Toshev~\cite{Toshev:1991ku}, which connect the Jarlskog invariants~\cite{Jarlskog:1985ht} in vacuum and in matter,  
\begin{eqnarray} 
&& 
\text{Naumov identity:} 
\hspace{4mm}
c_{23} s_{23} c^2_{13} s_{13} c_{12} s_{12} \sin \delta ~ 
( m^2_{2} - m^2_{1} ) ( m^2_{3} - m^2_{2} ) ( m^2_{3} - m^2_{1} ) 
\nonumber \\
&& 
\hspace{30mm}
=
\widetilde{ c_{23} } \widetilde{ s_{23} } \widetilde{ c_{13} }^2 \widetilde{ s_{13} } \widetilde{ c_{12} } \widetilde{ s_{12} }\sin \widetilde{ \delta }~
( \lambda_{2} - \lambda_{1} ) ( \lambda_{3} - \lambda_{2} )( \lambda_{3} - \lambda_{1} ), ~~
\label{Naumov}
\end{eqnarray}
\begin{eqnarray} 
&& 
\text{Toshev identity:} 
\hspace{6mm}
c_{23} s_{23} \sin \delta 
= \widetilde{ c_{23} } \widetilde{ s_{23} } \sin \widetilde{ \delta }, 
\label{Toshev}
\end{eqnarray}
where $\widetilde{ c_{23} }$ implies $\cos \widetilde{\theta_{23}}$, etc. Nature of the identity which originates from $H_{ \text{LHS} }=H_{ \text{RHS} }$ is transparent in the derivation of the Naumov identity in refs.~\cite{Harrison:1999df,Kimura:2002wd}.

If necessary for clarity, we can take the normal mass ordering and the neutrino channel in this section, though the extension to the alternative cases can be easily done. Then, $m^2_{1} < m^2_{2} < m^2_{3}$, and $\lambda_{1} < \lambda_{2} < \lambda_{3}$. In the vacuum limit $a \rightarrow 0$, $\lambda_{j} \rightarrow m^2_{j}$ ($j=1,2,3$) and $\widetilde{c_{ij}} \rightarrow c_{ij}$, $\widetilde{ \delta } \rightarrow \delta$ etc. Thanks to the above stated nature of the identities there is no ambiguity in the relative sign between the LHS and RHS of eqs.~\eqref{Naumov} and \eqref{Toshev}. 


\subsection{Toshev test for symmetry in the ZS system}
\label{sec:Toshev-ZS} 

Let us first examine the Toshev identity because it directly gives a powerful message on the relationship between the transformations of $H_{ \text{LHS} }$ and $H_{ \text{RHS} }$.
Under the transformations of Symmetry IA-vacuum in eq.~\eqref{Symmetry-IA-IB-vacuum}, the left-hand side (LHS) of eq.~\eqref{Toshev} does not transform, so that no transformation of the matter variables in the right-hand side (RHS) of eq.~\eqref{Toshev} is consistent. But, for Symmetry IB-vacuum in eq.~\eqref{Symmetry-IA-IB-vacuum}, the LHS of eq.~\eqref{Toshev} does transform by getting the minus sign. Therefore, at least one of the matter parameters in the RHS of eq.~\eqref{Toshev} must transform. 

Though the last statement above is in apparent contradiction to our previous statement ``the vacuum relabeling symmetry cannot affect the physical system'' in section~\ref{sec:matter-vs-vacuum}. However, since there is neither $\theta_{23}$ nor $\delta$ in the two-flavor model, there is no immediate inconsistency. The constraint from the Toshev identity seems to have a tension with the direct product structure of the vacuum and the matter symmetries advocated in ref.~\cite{Denton:2021vtf}.

\subsection{T-odd observables and the Naumov identity}
\label{sec:T-odd-obs} 

We point out that it is possible to make a nontrivial consistency check of the reparametrization symmetries of the state exchange type by using the T-odd observables in neutrino oscillation. The T-odd combination of the oscillation probabilities $\Delta P_{ \text{T} } \equiv P ( \nu_{\alpha} \rightarrow \nu_{\beta}) - P ( \nu_{\beta} \rightarrow \nu_{\alpha})$ in vacuum and in matter can be written as 
\begin{eqnarray} 
&& 
\Delta P_{ \text{Tvac} } 
=
16 c_{23} s_{23} c^2_{13} s_{13} c_{12} s_{12} \sin \delta ~ 
\sin \frac{ ( m^2_{2} - m^2_{1} ) x }{ 4E } 
\sin \frac{ ( m^2_{3} - m^2_{2} ) x }{ 4E } 
\sin \frac{ ( m^2_{3} - m^2_{1} ) x }{ 4E }, 
\nonumber \\
&&
\Delta P_{ \text{Tmatt} }
=
16 \widetilde{ c_{23} } \widetilde{ s_{23} } \widetilde{ c_{13} }^2 \widetilde{ s_{13} } \widetilde{ c_{12} } \widetilde{ s_{12} }\sin \widetilde{ \delta }~
\sin \frac{ ( \lambda_{2} - \lambda_{1} ) x }{ 4E } 
\sin \frac{ ( \lambda_{3} - \lambda_{2} ) x }{ 4E }
\sin \frac{ ( \lambda_{3} - \lambda_{1} ) x }{ 4E }. ~~
\nonumber \\
\label{T-odd-vacuum-matter}
\end{eqnarray}

We now discuss the response of $\Delta P_{ \text{Tvac} }$ and the LHS of the Naumov identity in eq.~\eqref{Naumov}. Under a state exchange symmetry transformation, the sign of $\Delta P_{ \text{Tvac} }$ may flip, but the same sign flip must occur in the LHS of the Naumov identity. Similarly, the sign flip in $\Delta P_{ \text{Tmatt} }$ is correlated to the sign flip in the RHS of the Naumov identity. That is, the sign responses of the Naumov identity are the physical observables. Therefore, one can make a consistency check of our state exchange symmetry using the Naumov identity. 

Let us make the Naumov test (or $\Delta P_{ \text{T} }$ test) of Symmetry IA- and IB-vacuum. The transformations of the former (latter) symmetry are given by $m^2_{1} \leftrightarrow m^2_{2}$, $c_{12} \rightarrow \mp s_{12}$, $s_{12} \rightarrow \pm c_{12}$ ($m^2_{1} \leftrightarrow m^2_{2}$, $c_{12} \leftrightarrow s_{12}$, $\delta \rightarrow \delta \pm \pi$). It is easy to verify that the LHS of the Naumov identity stays the same under these transformations. It is also readily verified that the RHS of the Naumov identity remains invariant under under the transformations of Symmetry IA- and IB-ZS \eqref{Symmetry-IA-IB-ZS}. 

\subsection{Naumov test for the DMP symmetries} 
\label{sec:Naumov-DMP}

We apply the Naumov test for the symmetries in DMP derived in ref.~\cite{Minakata:2021dqh}. Note that the Toshev test is trivial for them because $\theta_{23}$ and $\delta$ are not elevated to the matter variables. In our Naumov test we approximate the RHS of the Naumov identity in eq.~\eqref{Naumov} by the leading order expressions in the DMP perturbation theory, 
\begin{eqnarray} 
&& 
c_{23} s_{23} c^2_{13} s_{13} c_{12} s_{12} \sin \delta ~ 
( m^2_{2} - m^2_{1} ) ( m^2_{3} - m^2_{2} ) ( m^2_{3} - m^2_{1} ) 
\nonumber \\
&=& 
c_{23} s_{23} c^2_{\phi} s_{\phi} c_{\psi} s_{\psi} \sin \delta~
( \lambda_{2} - \lambda_{1} ) ( \lambda_{3} - \lambda_{2} )( \lambda_{3} - \lambda_{1} ).
\label{Naumov-DMP}
\end{eqnarray}
By making the leading order approximation, the Naumov identity becomes the approximate equality, and testing for the numerical accuracy in the both sides would not reveal a clear result. But, it is not the problem for us, because we confine ourselves to the sign test, as discussed above. That is, what we mean by the Naumov test for DMP is to verify the same, consistent sign non-flip in the both LHS and RHS of eq.~\eqref{Naumov-DMP} by all the DMP symmetries. 

One can show that for all the eight symmetries in DMP, Symmetry X-DMP, where X=IA, $\cdot \cdot \cdot $, IVB, listed in Table~1 in ref.~\cite{Minakata:2021dqh}, both the LHS and RHS of eq.~\eqref{Naumov-DMP} are invariant under the symmetry transformations. Therefore, all the DMP symmetries pass the Naumov sign test. Apart from Symmetry IA, the vacuum parameter transformations affect not only the LHS but also the RHS of eq.~\eqref{Naumov-DMP}, so that the results are highly nontrivial. It indicates a nontrivial consistency between the vacuum and the matter parameter transformations, the marked property of symmetries that are derived by the SF equation in matter. 

\section{Looking for the 1-3 state exchange symmetry in matter}
\label{sec:13-symmetry} 

The 1-3 state exchange symmetry takes a very special position among possible state-relabeling symmetries because it is practically impossible to construct in vacuum. However, we will show in this and the next sections that, with the SF method, it is indeed possible to formulate the 1-3 state exchange symmetry in matter. The flavor-mass eigenstate relation at the zeroth order, $\nu_{\alpha} = [ U_{23} (\theta_{23}) U_{13} (\phi, \delta) ]_{\alpha j} \hat{\nu}_{j}$ where $\hat{\nu}_{j}$ denotes the propagation basis (see eq.~\eqref{SF-eq-helioP-1st}), will be the key to allow us to find the 1-3 state exchange symmetry.

\subsection{Difficulty in constructing the 1-3 exchange symmetry in vacuum}
\label{sec:unique-13} 

With the conventional way of defining the flavor mixing matrix, $U=U_{23} (\theta_{23}) U_{13} (\theta_{13}, \delta) U_{12} (\theta_{12})$, it is hard to write down the pure $m^2_{1} \leftrightarrow m^2_{3}$ exchange without involving the other mass states. If one introduces the 1-3 state exchange in the $U_{13}$ rotation matrix, it has to pass through the $U_{12}$ rotation matrix to reach to the matter-mass eigenstate. Then, the $\nu_{2}$ state inevitably comes-in into the ``1-3'' state exchange.\footnote{
If we take a different $U$ matrix such as $U=U_{23} (\theta_{23}^{\prime}) U_{12} (\theta_{12}^{\prime}) U_{13} (\theta_{13}^{\prime}, \delta^{\prime} )$, see e.g., ref.~\cite{Denton:2020igp}, then one can do the job. But, in this case, $\theta_{ij}^{\prime}$ ($i,j=1,2,3$) and $\delta^{\prime}$ in this new parametrization is completely different from those in the our conventional parametrization. Rewriting the newly obtained 1-3 exchange symmetry transformations written with $\theta_{ij}^{\prime}$ ($i,j=1,2,3$) and $\delta^{\prime}$ by the three angles and CP phase in our PDG convention $U$ matrix would be a formidable task. 
}
The difficulty in constructing the 1-3 state exchange symmetry in vacuum and its availability in matter may open a window toward the better understanding of the state exchange symmetry in matter. 


\subsection{Renormalized helio-perturbation theory to first order}
\label{sec:helioP-1st}

We work with so called the renormalized helio-perturbation theory~\cite{Minakata:2015gra}, a particular version of the atmospheric-resonance perturbation theory~\cite{Arafune:1996bt,Cervera:2000kp,Freund:2001pn,Akhmedov:2004ny,Minakata:2015gra}, which describes the atmospheric-scale enhancement of neutrino oscillation~\cite{Mikheyev:1985zog,Barger:1980tf}, roughly phrased here as the ``resonance''~\cite{Smirnov:2016xzf}. 
The term ``helio'' is a shorthand of ``helio to terrestrial ratio'' which means the ratio of the solar to atmospheric (i.e., terrestrial) $\Delta m^2$. The renormalized helio-perturbation theory has the unique expansion parameter $\epsilon$ 
\begin{eqnarray} 
&&
\epsilon \equiv \frac{ \Delta m^2_{21} }{ \Delta m^2_{ \text{ren} } }, 
\hspace{10mm}
\Delta m^2_{ \text{ren} } \equiv \Delta m^2_{31} - s^2_{12} \Delta m^2_{21},
\label{epsilon-Dm2-ren-def}
\end{eqnarray}
where $\Delta m^2_{ \text{ren} }$ is the ``renormalized'' atmospheric $\Delta m^2$ used in ref.~\cite{Minakata:2015gra}. In what follows we simply call the theory as the ``helio-perturbation theory''. 

We follow the SF method~\cite{Minakata:2021dqh} to uncover the symmetry in the helio-perturbation theory. For convenience, we use the PDG convention~\cite{Zyla:2020zbs} of the flavor mixing matrix $U \equiv U_{\text{\tiny MNS}}$. Since the ATM convention $U$ matrix is used in ref.~\cite{Minakata:2015gra}, in which $e^{ \pm i \delta}$ is attached to $s_{23}$, we first transform everything into the PDG convention. Notice that the expression of the oscillation probability is identical independent of the conventions, ATM, PDG, and SOL, as the states are related only by the phase redefinition. 
The transformation of the flavor basis Hamiltonian and the flavor states from the ATM convention to the PDG can be done with 
\begin{eqnarray} 
H_{\text{\tiny PDG}} 
=
\left[
\begin{array}{ccc}
1 & 0 &  0  \\
0 & 1 & 0 \\
0 & 0 & e^{ i \delta} \\
\end{array}
\right] 
H_{\text{\tiny ATM}} 
\left[
\begin{array}{ccc}
1 & 0 &  0  \\
0 & 1 & 0 \\
0 & 0 & e^{- i \delta} \\
\end{array}
\right], 
\hspace{8mm} 
\nu_{\text{\tiny PDG}} 
= 
\left[
\begin{array}{ccc}
1 & 0 &  0  \\
0 & 1 & 0 \\
0 & 0 & e^{ i \delta} \\
\end{array}
\right] 
\nu_{\text{\tiny ATM}}.
\label{ATM-PDG}
\end{eqnarray}
See ref.~\cite{Martinez-Soler:2018lcy} for the relationships between the three $U$ matrix conventions.

The flavor basis state is expressed by the propagating eigenstate to first order in perturbation theory by using the $V$ matrix method~\cite{Minakata:1998bf}, with which the helio-perturbation theory is formulated~\cite{Minakata:2015gra}. In the PDG convention the relation can be written, using~\eqref{ATM-PDG}, as 
\begin{eqnarray} 
&& 
\left[
\begin{array}{c}
\nu_{e} \\
\nu_{\mu} \\
\nu_{\tau} \\
\end{array}
\right] 
= 
U_{23} (\theta_{23}) U_{13} (\phi, \delta) 
\biggl\{
1 + 
\epsilon c_{12} s_{12} 
\mathcal{ R } ( \phi, \delta; \lambda_{-}, \lambda_{+} ) 
\biggr\} 
\left[
\begin{array}{c}
\nu_{-} \\
\nu_{0} \\
\nu_{+} \\
\end{array}
\right], 
\label{SF-eq-helioP-1st}
\end{eqnarray}
where $\epsilon$ is the expansion parameter defined in eq.~\eqref{epsilon-Dm2-ren-def}, and $\mathcal{ R } ( \phi, \delta; \lambda_{-}, \lambda_{+} )$ is defined by 
\begin{eqnarray} 
&&
\mathcal{ R } ( \phi, \delta; \lambda_{-}, \lambda_{+} ) 
\equiv 
\left[ 
\begin{array}{ccc}
0 & -  c_{ \left( \phi - \theta_{13} \right) } 
\frac{ \Delta m^2_{ \text{ren} }  }{\lambda_{-} - \lambda_{0} } & 0 \\
 c_{ \left( \phi - \theta_{13} \right) } 
 \frac{ \Delta m^2_{ \text{ren} } }{\lambda_{-} - \lambda_{0} }  & 0 & s_{ \left( \phi - \theta_{13} \right) } e^{ - i \delta} \frac{ \Delta m^2_{ \text{ren} }  }{\lambda_{+} - \lambda_{0} } \\
0 & -  s_{ \left( \phi - \theta_{13} \right) } e^{ i \delta} 
\frac{ \Delta m^2_{ \text{ren} } }{\lambda_{+} - \lambda_{0} } & 0
\end{array}
\right], 
\label{mathcal-R}
\end{eqnarray}
where $c_{ \left( \phi - \theta_{13} \right) } \equiv \cos \left( \phi - \theta_{13} \right)$, and $s_{ \left( \phi - \theta_{13} \right) } \equiv \sin \left( \phi - \theta_{13} \right)$ are the abbreviated notations. 

Our state label [$\nu_{-}$, $\nu_{0}$, $\nu_{+}$] is arranged in such a way that $\nu_{+}$ and $\nu_{-}$ always undergo the atmospheric level crossing. The relationship between our state label [$\nu_{-}$, $\nu_{0}$, $\nu_{+}$] and the standard notation [$\nu_{1}$, $\nu_{2}$, $\nu_{3}$] (with the property $\lambda_{1} < \lambda_{2} <\lambda_{3}$ in the normal mass ordering) is that [$\nu_{1}$, $\nu_{2}$, $\nu_{3}$] correspond to [$\nu_{0}$, $\nu_{-}$, $\nu_{+}$] above the solar level crossing, and to [$\nu_{-}$ $\nu_{0}$ $\nu_{+}$] below the solar level crossing in both the normal and the inverted mass orderings. By ``above the solar level crossing'' we mean by $\rho E$ so that the region $\rho E < 0$ corresponds to the antineutrino channels. See Fig.~3 in ref.~\cite{Minakata:2015gra}. 

\subsection{Symmetry Finder (SF) equation in the helio-perturbation theory} 
\label{sec:SF-eq-helioP}

Following the spirit of eq.~\eqref{flavor-mass-vacuum} in the vacuum case, we look for an alternative form of eq.~\eqref{SF-eq-helioP-1st} in which the transformed $V$ matrix acts on the $\nu_{-} - \nu_{+}$ exchanged state. It naturally leads us to the ansatz 
\begin{eqnarray} 
&& 
F 
\left[
\begin{array}{c}
\nu_{e} \\
\nu_{\mu} \\
\nu_{\tau} \\
\end{array}
\right] 
= 
F U_{23} (\theta_{23}) U_{13} (\phi, \delta) 
G^{\dagger} G 
\biggl\{
1 + 
\epsilon c_{12} s_{12} 
\mathcal{ R } ( \phi, \delta; \lambda_{-}, \lambda_{+} ) 
\biggr\} 
G^{\dagger} G
\left[
\begin{array}{c}
\nu_{-} \\
\nu_{0} \\
\nu_{+} \\
\end{array}
\right].
\label{SF-eq-helioP-ansatz}
\end{eqnarray}
In eq.~\eqref{SF-eq-helioP-ansatz} we have introduced the flavor-state rephasing matrix $F$, which is parametrized as 
\begin{eqnarray} 
&&
F \equiv 
\left[
\begin{array}{ccc}
e^{ i \tau } & 0 & 0 \\
0 & 1 & 0 \\
0 & 0 & e^{ i \sigma } \\
\end{array}
\right],
\label{F-def}
\end{eqnarray}
and a generalized $\nu_{-} \leftrightarrow \nu_{+}$ state exchange matrix $G$
\begin{eqnarray} 
&& 
G \equiv 
\left[
\begin{array}{ccc}
0 & 0 & - e^{ - i ( \delta - \alpha) } \\
0 & 1 & 0 \\
e^{ i ( \delta - \beta) } & 0 & 0 \\
\end{array}
\right], 
\hspace{8mm}
G^{\dagger} \equiv 
\left[
\begin{array}{ccc}
0 & 0 & e^{ - i ( \delta - \beta) } \\
0 & 1 & 0 \\
- e^{ i ( \delta - \alpha) } & 0 & 0 \\
\end{array}
\right],
\label{G-def}
\end{eqnarray}
where $\tau$, $\sigma$, $\alpha$, and $\beta$ denote the arbitrary phases. Notice that the rephasing and state exchange matrices, $F$ and $G$ in eqs.~\eqref{F-def} and \eqref{G-def}, takes the nonvanishing, nontrivial (not unity) elements in $\nu_{-} - \nu_{+}$ sub-sector. It is because we restrict ourselves into the $\nu_{-} \leftrightarrow \nu_{+}$ state exchange symmetry. 

The SF equation, an explicit form of eq.~\eqref{SF-eq-helioP-ansatz}, reads 
\begin{eqnarray} 
&&
\hspace{-10mm} 
\left[
\begin{array}{ccc}
e^{ i \tau } & 0 & 0 \\
0 & 1 & 0 \\
0 & 0 & e^{ i \sigma } \\
\end{array}
\right]
\left[
\begin{array}{c}
\nu_{e} \\
\nu_{\mu} \\
\nu_{\tau} \\
\end{array}
\right] 
=  
\left[
\begin{array}{ccc}
1 & 0 &  0  \\
0 & c_{23} & s_{23} e^{ - i \sigma } \\
0 & - s_{23} e^{ i \sigma } & c_{23} \\
\end{array}
\right] 
F U_{13} (\phi, \delta) G^{\dagger} 
G 
\biggl\{
1 + 
\epsilon c_{12} s_{12} 
\mathcal{ R } ( \phi, \delta; \lambda_{-}, \lambda_{+} ) 
\biggr\} 
G^{\dagger}
G 
\left[
\begin{array}{c}
\nu_{1} \\
\nu_{2} \\
\nu_{3} \\
\end{array}
\right] 
\nonumber \\
&=& 
\left[
\begin{array}{ccc}
1 & 0 &  0  \\
0 & c_{23}^{\prime} & s_{23}^{\prime} \\
0 & - s_{23}^{\prime} & c_{23}^{\prime} \\
\end{array}
\right] 
U_{13} ( \phi^{\prime}, \delta + \xi) 
\biggl\{
1 + 
\epsilon c_{12}^{\prime} s_{12}^{\prime} 
\mathcal{R} ( \phi^{\prime}, \delta + \xi; \lambda_{+}, \lambda_{-} ) 
\biggr\} 
\left[
\begin{array}{c}
- e^{ - i ( \delta - \alpha) } \nu_{3} \\
\nu_{2} \\
e^{ i ( \delta - \beta) } \nu_{1} \\
\end{array}
\right].
\label{SF-eq-helioP}
\end{eqnarray}
We would like to keep the transformed $s_{23}$, $s_{23}^{\prime} = s_{23} e^{ - i \sigma }$, a real number, which means that $\sigma$ must be an integral multiple of $\pi$. Under this ansatz, the SF equation~\eqref{SF-eq-helioP} can be decomposed into the following first and the second conditions. They read 
\begin{eqnarray} 
&&
F U_{13} (\phi, \delta) G^{\dagger} = U_{13} ( \phi^{\prime}, \delta + \xi), 
\nonumber \\
&& 
\epsilon c_{12} s_{12} G 
\mathcal{ R } ( \phi, \delta; \lambda_{-}, \lambda_{+} ) 
G^{\dagger} 
= 
\epsilon c_{12}^{\prime} s_{12}^{\prime} 
\mathcal{R} ( \phi^{\prime}, \delta + \xi; \lambda_{+}, \lambda_{-} ) 
\label{SF-eq-1st-2nd}
\end{eqnarray}
The explicit form of the first condition is given by 
\begin{eqnarray} 
&&
\left[
\begin{array}{ccc}
- s_{\phi} e^{ - i ( \alpha - \tau ) }  & 0 & c_{\phi} e^{ - i ( \delta - \beta - \tau ) } \\
0 & 1 & 0 \\
- c_{\phi} e^{ i ( \delta - \alpha + \sigma ) } & 0 & - s_{\phi} e^{ i ( \beta + \sigma ) } \\
\end{array}
\right] 
=
\left[
\begin{array}{ccc}
c_{\phi}^{\prime} & 0 & s_{\phi}^{\prime} e^{- i ( \delta + \xi) } \\
0 & 1 & 0 \\
- s_{\phi}^{\prime} e^{ i ( \delta + \xi) } & 0 & c_{\phi}^{\prime} \\
\end{array}
\right],
\label{1st-condition-explicit}
\end{eqnarray}
and the second condition by 
\begin{eqnarray} 
&&
\epsilon c_{12} s_{12} 
\left[ 
\begin{array}{ccc}
0 & s_{ \left( \phi - \theta_{13} \right) } e^{ i \alpha } \frac{ \Delta m^2_{ \text{ren} } }{\lambda_{+} - \lambda_{0} } 
& 0 \\
- s_{ \left( \phi - \theta_{13} \right) } e^{ - i \alpha } 
\frac{ \Delta m^2_{ \text{ren} }  }{\lambda_{+} - \lambda_{0} } & 
0 & c_{ \left( \phi - \theta_{13} \right) } e^{ - i ( \delta - \beta) } \frac{ \Delta m^2_{ \text{ren} } }{\lambda_{-} - \lambda_{0} } \\
0 & 
- c_{ \left( \phi - \theta_{13} \right) } e^{ i ( \delta - \beta) } \frac{ \Delta m^2_{ \text{ren} }  }{\lambda_{-} - \lambda_{0} } & 
0
\end{array}
\right]  
\nonumber \\
&=& 
\epsilon c_{12}^{\prime} s_{12}^{\prime} 
\left[ 
\begin{array}{ccc}
0 & - c_{ ( \phi^{\prime} - \theta_{13}^{\prime} ) } 
\frac{ \Delta m^2_{ \text{ren} }  }{ \lambda_{+} - \lambda_{0} } & 0 \\
 c_{ ( \phi^{\prime} - \theta_{13}^{\prime} ) } 
 \frac{ \Delta m^2_{ \text{ren} } }{\lambda_{+} - \lambda_{0} }  & 0 & 
 s_{ ( \phi^{\prime} - \theta_{13}^{\prime} ) } e^{ - i ( \delta + \xi ) } \frac{ \Delta m^2_{ \text{ren} }  }{\lambda_{-} - \lambda_{0} } \\
0 & - s_{ ( \phi^{\prime} - \theta_{13}^{\prime} ) } e^{ i ( \delta + \xi ) } 
\frac{ \Delta m^2_{ \text{ren} } }{\lambda_{-} - \lambda_{0} } & 0
\end{array}
\right], 
\label{2nd-condition-full} 
\end{eqnarray}
where the notation is such that $c_{12}^{\prime} \equiv \cos \theta_{12}^{\prime}$, and $c_{ ( \phi^{\prime} - \theta_{13}^{\prime} ) } \equiv \cos ( \phi^{\prime} - \theta_{13}^{\prime} )$ etc.

Here is an important note for $\tau$, $\sigma$, $\alpha$, $\beta$, and $\xi$. We have already stated above that $\sigma$ in units of $\pi$ is an integer to keep $s_{23}$ a real number. Similarly, eq~\eqref{1st-condition-explicit} tells us that $\beta + \sigma$ and $\alpha - \tau$ must be integers, where we abbreviate ``in units of $\pi$'' for the moment. Then, $\beta$ must be an integer as well. Now, the second condition~\eqref{2nd-condition-full} requires that $\alpha$ must be an integer, which implies that $\tau$ must be an integer. Compare the 1-2 or 2-1 elements of the both sides. Then, by comparing the 2-3 elements at the both sides we know that $\xi$ is an integer. Thus, we have shown that $\tau$, $\sigma$, $\alpha$, $\beta$, and $\xi$ are all integers in units of $\pi$.

Though our SF equation~\eqref{SF-eq-1st-2nd} is similar to that in DMP~\cite{Minakata:2021dqh}, there is an important difference between them. In the present system, $\mathcal{ R } ( \phi, \delta; \lambda_{-}, \lambda_{+} )$ in eq.~\eqref{SF-eq-1st-2nd}  contains the both $\phi$ and $\theta_{13}$, in the particular combination $\phi - \theta_{13}$. In DMP, the similar $\mathcal{ R }$ matrix does not contain the vacuum mixing angles. The difference entails the doubled, sixteen symmetries in the helio-perturbation theory.

\subsection{Analyzing the first condition}
\label{sec:1st-condition}

We analyze the first condition~\eqref{1st-condition-explicit}. It is not difficult to show that it entails the conditions 
\begin{eqnarray} 
&&
c_{\phi^{\prime}} 
= - s_{\phi} e^{ - i ( \alpha - \tau ) } 
= - s_{\phi} e^{ i ( \beta + \sigma ) }, 
\hspace{8mm}
s_{\phi^{\prime}} 
= c_{\phi} e^{ i ( \beta + \tau + \xi ) } 
= c_{\phi} e^{ - i ( \alpha - \sigma + \xi ) }, 
\label{1st-condition-helioP}
\end{eqnarray}
and the consistency conditions for the phases 
\begin{eqnarray} 
&&
\alpha + \beta - \tau + \sigma = 0 ~~~(\text{mod.} ~2\pi), 
\hspace{8mm}
\tau - \sigma + \xi = 0, ~\pm \pi
\label{1st-sol-consistency} 
\end{eqnarray}
In the second equation in eq.~\eqref{1st-sol-consistency}, a classification naturally appeared: 
\begin{eqnarray} 
&&
\text{ Class I: } \tau - \sigma = - \xi, 
\hspace{8mm}
\text{ Class II: } \tau - \sigma = - \xi \pm \pi.
\label{Class-I-II}
\end{eqnarray}

Then, the procedure for obtaining solutions to the SF equation is: 
(1) To choose an ansatz for $\xi$. In this paper we try only the limited choices, $\xi = 0, \pi$. (2) To choose Class I or Class II. Then, find all possible solutions for $\tau$, $\sigma$, $\alpha$, and $\beta$. (3) Verify the solution against the second condition~\eqref{2nd-condition-full}. 

\section{The 1-3 state exchange symmetry: Analysis and result}
\label{sec:13-symmetry-result}

Let us present a few examples of the symmetries as the solutions to the SF equation.  
We start our search from the easiest case of no flavor-state rephasing, $\tau = \sigma = 0$. In the following discussion, we mean by Symmetry Type A a symmetry whose transformations do not include $\delta$, and Type B a symmetry which includes transformation of $\delta$. In the latter we only consider a shift of $\delta$ with the amount $\pm \pi$. This $\pm$ sign is not important, as $\delta$ is a periodic variable with period $2\pi$. We denote the case of no favor-basis rephasing as Type I, so that we will be considering Symmetry IA and IB in the following two sections before section~\ref{sec:whole-structure}. 

\subsection{Symmetry IA in the helio-perturbation theory} 
\label{sec:IA-helioP}

We investigate the simplest solution $\tau = \sigma = \xi = 0$. There are two solutions of eq.~\eqref{1st-condition-helioP}, $\alpha = \beta = 0$ (upper sign) and $\alpha = \pi$, and $\beta = - \pi$ (lower sign). This is in Class I. 
Because we want to observe more clearly the new feature of doubled solutions due to $s_{12}$ sign flip - nonflip dualism, we treat the upper-sign and lower-sign cases separately. 

\vspace{1mm}

\noindent 
{\bf Case of [$\tau = 0, \sigma = 0$, $\xi = 0$ $\alpha = 0, \beta = 0$]}: 

The solution of the first condition eq.~\eqref{1st-condition-helioP} is given by 
$c_{\phi^{\prime}} = - s_{\phi}$, 
$s_{\phi^{\prime}} = c_{\phi}$, which implies 
$\phi^{\prime} = \phi + \frac{\pi}{2}$. 
The solution to the second condition~\eqref{2nd-condition-full} is given by 
\begin{eqnarray} 
&& 
\text{ (a) No sign flip of $s_{12}$: } 
\hspace{6mm} 
c _{ ( \phi^{\prime} - \theta_{13}^{\prime} ) } = - s_{ ( \phi - \theta_{13} ) }, 
\hspace{6mm} 
s _{ ( \phi^{\prime} - \theta_{13}^{\prime} ) } = c_{ ( \phi - \theta_{13} ) }, 
\nonumber \\
&& 
\text{ (b) With sign flip of $s_{12}$: } 
\hspace{6mm} 
c _{ ( \phi^{\prime} - \theta_{13}^{\prime} ) } = s_{ ( \phi - \theta_{13} ) }, 
\hspace{6mm} 
s _{ ( \phi^{\prime} - \theta_{13}^{\prime} ) } = - c_{ ( \phi - \theta_{13} ) }.
\label{s12-flip-noflip-IA}
\end{eqnarray}
One might think that one of them must be rejected as a solution, as in the case of DMP SF equation. Here, the situation is more complicated, and we examine one by one of these two cases. 

\noindent
[No sign flip of $s_{12}$]: 
The solution in eq.~\eqref{s12-flip-noflip-IA} implies that 
$( \phi^{\prime} - \theta_{13}^{\prime} ) = ( \phi - \theta_{13} ) + \frac{\pi}{2}$. Together with  $\phi^{\prime} = \phi + \frac{\pi}{2}$ above,\footnote{
We still keep our attitude to remain the region of definition of the mixing angles $0 \leq \phi \leq \frac{\pi}{2}$, but for simplicity of notation we use the expression such as $\phi^{\prime} = \phi + \frac{\pi}{2}$. A precise description of the prescription which we impose to embody our attitude is given in section 3.2 in ref.~\cite{Minakata:2021dqh}. 
}
the resultant relation is $\theta_{13}^{\prime} = \theta_{13}$. That is, $\theta_{13}$ does not transform. 

\noindent
[With sign flip of $s_{12}$]: 
The solution in eq.~\eqref{s12-flip-noflip-IA} with sign flip of $s_{12}$ implies that 
$( \phi^{\prime} - \theta_{13}^{\prime} ) = ( \phi - \theta_{13} ) - \frac{\pi}{2}$. When combined with $\phi^{\prime} = \phi + \frac{\pi}{2}$, the solution is given by 
$\theta_{13}^{\prime} = \theta_{13} + \pi$. 

To convince ourselves that the both solutions are tenable, we appeal to the expressions of $\cos 2 \phi$ and $\sin 2 \phi$~\cite{Minakata:2015gra}:
\begin{eqnarray} 
\cos 2 \phi 
&=& 
\frac{ \Delta m^2_{ \text{ren} } \cos 2\theta_{13} - a }{ \lambda_{+} - \lambda_{-} },
\nonumber \\
\sin 2 \phi 
&=& 
\frac{ \Delta m^2_{ \text{ren} } \sin 2\theta_{13} }{ \lambda_{+} - \lambda_{-} }.
\label{cos-sin-2phi}
\end{eqnarray}
In the case of [No sign flip of $s_{12}$], $\theta_{13}$ does not transform. But since our transformation involve the state exchange $\lambda_{-} \leftrightarrow \lambda_{+}$, the transformation of $\phi$ is such that 
$\cos 2 \phi \rightarrow - \cos 2 \phi$ and $\sin 2 \phi \rightarrow - \sin 2 \phi$. 
They are consistent with $2 \phi^{\prime} = 2 \phi + \pi$, as above. Therefore, we have shown that [No sign flip of $s_{12}$] solution is a consistent solution. 
Now, let us discuss [With sign flip of $s_{12}$] solution in which case $\theta_{13}^{\prime} = \theta_{13} + \pi$. Even though $\theta_{13}$ transforms, because $2 \theta_{13}^{\prime} = 2 \theta_{13}$ mod. $2\pi$, the transformation of $\phi$ is the same as above. Therefore, the solution [With sign flip of $s_{12}$] also qualifies as a viable solution. 

We would like to stress here again that unlike the case of DMP symmetries, a pair of flip and non-flip $s_{12}$ solutions are always allowed. In DMP, one of these options must be chosen as dictated by the SF equation. The both options become available in our present case because of the cooperation of $\theta_{13}$ which now lives in the $\mathcal{R}$ matrix. That is, possible inconsistency in one of the two solutions is taken care of by the change in $\theta_{13}$, in a way consistent with the $\phi$ transformation. We will see this feature in all the rest of the symmetries. 

\vspace{1mm}

\noindent 
{\bf Case of [$\tau = 0, \sigma = 0$, $\xi = 0$ $\alpha = \pi, \beta = - \pi$]}: 

We now turn to the lower sign solution. The solution of the first condition eq.~\eqref{1st-condition-helioP} is given by 
$c_{\phi^{\prime}} = s_{\phi}$, 
$s_{\phi^{\prime}} = - c_{\phi}$, which implies 
$\phi^{\prime} = \phi - \frac{\pi}{2}$. 
The solution to the second condition~\eqref{2nd-condition-full} is given by 
\begin{eqnarray} 
&& 
\text{ (a) No sign flip of $s_{12}$: } 
\hspace{6mm} 
c _{ ( \phi^{\prime} - \theta_{13}^{\prime} ) } = s_{ ( \phi - \theta_{13} ) }, 
\hspace{6mm} 
s _{ ( \phi^{\prime} - \theta_{13}^{\prime} ) } = - c_{ ( \phi - \theta_{13} ) }, 
\nonumber \\
&& 
\text{ (b) With sign flip of $s_{12}$: } 
\hspace{6mm} 
c _{ ( \phi^{\prime} - \theta_{13}^{\prime} ) } = - s_{ ( \phi - \theta_{13} ) }, 
\hspace{6mm} 
s _{ ( \phi^{\prime} - \theta_{13}^{\prime} ) } = c_{ ( \phi - \theta_{13} ) }.
\label{s12-flip-noflip2}
\end{eqnarray}
We can proceed as in the case of the upper sign. The transformation property of $\theta_{13}$ is given by 
$\theta_{13}^{\prime} = \theta_{13}$ in [No sign flip of $s_{12}$], 
$\theta_{13}^{\prime} = \theta_{13} - \pi$ in [With sign flip of $s_{12}$]. 
The both solutions qualify as the consistent solutions of the SF equation. 

It is important to confirm that the symmetry transformations we have obtained do really exchange the eigenvalues, $\lambda_{-} \leftrightarrow \lambda_{+}$, when we perform the transformations. It can be done easily by the alternative expressions of the eigenvalues 
\begin{eqnarray} 
\lambda_{-} &=& 
\sin^2 \left( \phi - \theta_{13} \right) \Delta m^2_{ \text{ren} } 
+ c^2_{\phi} a  
+ \epsilon s^2_{12} \Delta m^2_{ \text{ren} },
\nonumber \\
\lambda_{+} &=&
\cos^2 \left( \phi - \theta_{13} \right) \Delta m^2_{ \text{ren} } 
+ s^2_{\phi} a  
+ \epsilon s^2_{12} \Delta m^2_{ \text{ren} }.
\label{eigenvalues}
\end{eqnarray}
The transformations we have obtained above are of the two types, exchanging between $c_{\phi}$ and $s_{\phi}$, and exchanging between $c _{ ( \phi^{\prime} - \theta_{13}^{\prime} ) }$ and $s_{ ( \phi - \theta_{13} ) }$, but they are done simultaneously. Then, given the expressions of the eigenvalues in eq.~\eqref{eigenvalues}, the $\phi$ transformations are consistent with the eigenvalue exchange. We did the consistency check here for IA type symmetries in the helio-perturbation theory, but it will be repeated in all the other symmetries, even though we do not mention it explicitly. 

\subsubsection{Symmetry IA-helioP and Symmetry IAf-helioP} 
\label{sec:IA-IAf-helioP}

Thus, we have obtained the following two sets of the solutions, ``Symmetry IA-helioP'' and ``Symmetry IAf-helioP'', where $``f''$ denotes the sign flip of $s_{12}$:
\begin{eqnarray} 
&& 
\hspace{-6mm}
\text{Symmetry IA-helioP:}
\hspace{8mm}
\lambda_{-} \leftrightarrow \lambda_{+}, 
\hspace{6mm}
c_{\phi} \rightarrow \mp s_{\phi}, 
\hspace{6mm}
s_{\phi} \rightarrow \pm c_{\phi}, 
\hspace{6mm}
\phi \rightarrow \phi \pm \frac{\pi}{2}, 
\nonumber \\
&& 
\hspace{38mm}
 c_{ \left( \phi - \theta_{13} \right) } 
 \rightarrow 
 \mp s_{ \left( \phi - \theta_{13} \right) }, 
 \hspace{6mm}
 s_{ \left( \phi - \theta_{13} \right) } 
 \rightarrow 
 \pm c_{ \left( \phi - \theta_{13} \right) }, 
\nonumber \\
&& 
\hspace{38mm}
 c_{\phi} c_{ \left( \phi - \theta_{13} \right) } 
 \rightarrow 
 s_{\phi} s_{ \left( \phi - \theta_{13} \right) },  
 \hspace{6mm}
 s_{\phi} s_{ \left( \phi - \theta_{13} \right) } 
 \rightarrow 
 c_{\phi} c_{ \left( \phi - \theta_{13} \right) }. 
\label{Symmetry-IA-helioP}
\end{eqnarray}
\begin{eqnarray} 
&& 
\hspace{-4mm}
\text{Symmetry IAf-helioP:} 
\hspace{8mm}
\theta_{13} \rightarrow \theta_{13} \pm \pi, 
\hspace{8mm}
\theta_{12} \rightarrow - \theta_{12}, 
\nonumber \\
&& 
\hspace{38mm}
\lambda_{-} \leftrightarrow \lambda_{+}, 
\hspace{6mm}
c_{\phi} \rightarrow \mp s_{\phi}, 
\hspace{6mm}
s_{\phi} \rightarrow \pm c_{\phi}, 
\hspace{6mm}
\phi \rightarrow \phi \pm \frac{\pi}{2}, 
\nonumber \\
&& 
\hspace{38mm}
 c_{ \left( \phi - \theta_{13} \right) } 
 \rightarrow 
 \pm s_{ \left( \phi - \theta_{13} \right) }, 
 \hspace{6mm}
 s_{ \left( \phi - \theta_{13} \right) } 
 \rightarrow 
 \mp c_{ \left( \phi - \theta_{13} \right) }, 
\nonumber \\
&& 
\hspace{38mm}
 c_{\phi} c_{ \left( \phi - \theta_{13} \right) } 
 \rightarrow 
- s_{\phi} s_{ \left( \phi - \theta_{13} \right) },  
 \hspace{6mm}
 s_{\phi} s_{ \left( \phi - \theta_{13} \right) } 
 \rightarrow 
- c_{\phi} c_{ \left( \phi - \theta_{13} \right) }. 
\label{Symmetry-IAf-helioP}
\end{eqnarray}
The last lines in eqs.~\eqref{Symmetry-IA-helioP} and \eqref{Symmetry-IAf-helioP} may be redundant, but they are given for convenience of the readers to verify that the oscillation probability is invariant under the transformations of these symmetries. Now, it should be straightforward to verify Symmetry IA-helioP and Symmetry IAf-helioP against the expressions of the oscillation probability given in Appendix~B in ref.~\cite{Minakata:2015gra}. 

Notice that in Symmetry IA-helioP no transformation of the vacuum parameter is involved. It is nothing but the unique symmetry observed prior to this work in ref.~\cite{Martinez-Soler:2019nhb}, the symmetry of 1-3 state exchange type uncovered for the first time to the best of our knowledge. In this paper, we will learn that it is just one of the sixteen. 

\subsection{Symmetry IB in the helio-perturbation theory} 
\label{sec:IB-helioP}

After learning lessons with Symmetry IA it is a straightforward exercise to discuss the 
Symmetry IB with the $\delta \rightarrow \delta + \pi$ transformation. We have the following two cases: 
[$\tau = 0, \sigma = 0$, $\xi = \pi$ $\alpha = \pi, \beta = - \pi$] (upper sign), and 
[$\tau = 0, \sigma = 0$, $\xi = \pi$ $\alpha = 0, \beta = 0$] (lower sign). 
They are in Class II. 

\vspace{1mm}

\noindent 
{\bf Case of [$\tau = 0, \sigma = 0$, $\xi = \pi$ $\alpha = \pi, \beta = - \pi$]}

\vspace{1mm}

The solution of the first condition eq.~\eqref{1st-condition-helioP} is given by 
$c_{\phi^{\prime}} = s_{\phi}$, 
$s_{\phi^{\prime}} = c_{\phi}$, which implies 
$\phi^{\prime} = - \phi + \frac{\pi}{2}$. 
The solution to the second condition~\eqref{2nd-condition-full} is given by 
\begin{eqnarray} 
&& 
\text{ (a) No sign flip of $s_{12}$: } 
\hspace{6mm} 
c _{ ( \phi^{\prime} - \theta_{13}^{\prime} ) } = s_{ ( \phi - \theta_{13} ) }, 
\hspace{6mm} 
s _{ ( \phi^{\prime} - \theta_{13}^{\prime} ) } = c_{ ( \phi - \theta_{13} ) }, 
\nonumber \\
&& 
\text{ (b) With sign flip of $s_{12}$: } 
\hspace{6mm} 
c _{ ( \phi^{\prime} - \theta_{13}^{\prime} ) } = - s_{ ( \phi - \theta_{13} ) }, 
\hspace{6mm} 
s _{ ( \phi^{\prime} - \theta_{13}^{\prime} ) } = - c_{ ( \phi - \theta_{13} ) }.
\label{s12-flip-noflip-IB}
\end{eqnarray}
The transformations of $( \phi - \theta_{13} )$ in the above two cases are: 
(a) $\phi^{\prime} - \theta_{13}^{\prime} = - ( \phi - \theta_{13} ) + \frac{\pi}{2}$, and 
(b) $\phi^{\prime} - \theta_{13}^{\prime} = - ( \phi - \theta_{13} ) - \frac{\pi}{2}$, which lead to the $\theta_{13}$ transformation properties of 
(a) $\theta_{13}^{\prime} = - \theta_{13}$, and 
(b) $\theta_{13}^{\prime} = - \theta_{13} + \pi$, respectively. 
Using eq.~\eqref{cos-sin-2phi}, they both implies that $\sin 2\phi^{\prime} = \sin 2\phi$ and $\cos 2\phi^{\prime} = - \cos 2\phi$, which is consistent with the transformation property of $\phi$ which comes from the first condition, $\phi^{\prime} = - \phi \pm \frac{\pi}{2}$.

\vspace{1mm}

\noindent 
{\bf Case of [$\tau = 0, \sigma = 0$, $\xi = \pi$ $\alpha = 0, \beta = 0$]}

\vspace{1mm}

The treatment of the lower sign case is as before with the solution of the first condition eq.~\eqref{1st-condition-helioP} given by  
$c_{\phi^{\prime}} = - s_{\phi}$, 
$s_{\phi^{\prime}} = - c_{\phi}$, which implies 
$\phi^{\prime} = - \phi - \frac{\pi}{2}$. 
The solution to the second condition~\eqref{2nd-condition-full} is given by 
\begin{eqnarray} 
&& 
\text{ (a) No sign flip of $s_{12}$: } 
\hspace{6mm} 
c _{ ( \phi^{\prime} - \theta_{13}^{\prime} ) } = - s_{ ( \phi - \theta_{13} ) }, 
\hspace{6mm} 
s _{ ( \phi^{\prime} - \theta_{13}^{\prime} ) } = - c_{ ( \phi - \theta_{13} ) }, 
\nonumber \\
&& 
\text{ (b) With sign flip of $s_{12}$: } 
\hspace{6mm} 
c _{ ( \phi^{\prime} - \theta_{13}^{\prime} ) } = s_{ ( \phi - \theta_{13} ) }, 
\hspace{6mm} 
s _{ ( \phi^{\prime} - \theta_{13}^{\prime} ) } = c_{ ( \phi - \theta_{13} ) }.
\label{s12-flip-noflip2}
\end{eqnarray}
The transformation properties of $\theta_{13}$ is such that 
(a) $\theta_{13}^{\prime} = - \theta_{13}$, and 
(b) $\theta_{13}^{\prime} = - \theta_{13} - \pi$, respectively. 
The consistency between the $\theta_{13}$ and the $\phi$ transformation properties can be checked in exactly the same way as above. 

\subsubsection{Symmetry IB-helioP and Symmetry IBf-helioP} 
\label{sec:IB-IBf-helioP}

Thus, we have obtained the another two sets of the solutions, 
``Symmetry IB-helioP'' and ``Symmetry IBf-helioP'', where $``f''$ denotes the sign flip of $s_{12}$:
\begin{eqnarray} 
&& 
\hspace{-4mm}
\text{Symmetry IB-helioP:}
\hspace{8mm} 
\theta_{13} \rightarrow - \theta_{13}, 
\hspace{8mm} 
\delta \rightarrow \delta + \pi,
\nonumber \\
&& 
\hspace{40mm}
\lambda_{-} \leftrightarrow \lambda_{+}, 
\hspace{6mm}
c_{\phi} \rightarrow \pm s_{\phi}, 
\hspace{6mm}
s_{\phi} \rightarrow \pm c_{\phi}, 
\hspace{6mm}
\phi \rightarrow - \phi \pm \frac{\pi}{2}, 
\nonumber \\
&& 
\hspace{40mm}
 c_{ \left( \phi - \theta_{13} \right) } 
 \rightarrow 
 \pm s_{ \left( \phi - \theta_{13} \right) }, 
 \hspace{6mm}
 s_{ \left( \phi - \theta_{13} \right) } 
 \rightarrow 
 \pm c_{ \left( \phi - \theta_{13} \right) }, 
\nonumber \\
&& 
\hspace{40mm}
 c_{\phi} c_{ \left( \phi - \theta_{13} \right) } 
 \rightarrow 
 s_{\phi} s_{ \left( \phi - \theta_{13} \right) },  
 \hspace{6mm}
 s_{\phi} s_{ \left( \phi - \theta_{13} \right) } 
 \rightarrow 
 c_{\phi} c_{ \left( \phi - \theta_{13} \right) }. 
\label{Symmetry-IB-helioP}
\end{eqnarray}
\begin{eqnarray} 
&& 
\hspace{-5mm}
\text{Symmetry IBf-helioP:} 
\hspace{8mm}
\theta_{13} \rightarrow - \theta_{13} \pm \pi, 
\hspace{8mm}
\theta_{12} \rightarrow - \theta_{12}, 
\hspace{8mm} 
\delta \rightarrow \delta + \pi,
\nonumber \\
&& 
\hspace{38mm}
\lambda_{-} \leftrightarrow \lambda_{+}, 
\hspace{6mm}
c_{\phi} \rightarrow \pm s_{\phi}, 
\hspace{6mm}
s_{\phi} \rightarrow \pm c_{\phi}, 
\hspace{6mm}
\phi \rightarrow - \phi \pm \frac{\pi}{2}, 
\nonumber \\
&& 
\hspace{38mm}
 c_{ \left( \phi - \theta_{13} \right) } 
 \rightarrow 
 \mp s_{ \left( \phi - \theta_{13} \right) }, 
 \hspace{6mm}
 s_{ \left( \phi - \theta_{13} \right) } 
 \rightarrow 
 \mp c_{ \left( \phi - \theta_{13} \right) }, 
\nonumber \\
&& 
\hspace{38mm}
 c_{\phi} c_{ \left( \phi - \theta_{13} \right) } 
 \rightarrow 
- s_{\phi} s_{ \left( \phi - \theta_{13} \right) },  
 \hspace{6mm}
 s_{\phi} s_{ \left( \phi - \theta_{13} \right) } 
 \rightarrow 
- c_{\phi} c_{ \left( \phi - \theta_{13} \right) }. 
\label{Symmetry-IBf-helioP}
\end{eqnarray}
Again the above transformations of Symmetry IB-helioP and Symmetry IBf-helioP leave the oscillation probabilities to first order in helio-perturbation theory invariant. It is surprising that such a rich symmetry structure was hidden, remained undetected in the previous investigations. 

\subsection{The whole structure of symmetry in the helio-perturbation theory }
\label{sec:whole-structure}

The similar job of solving the SF equation can be repeated for the other solutions of the parameters that appear in the equation. Since the computation is so similar that we hesitate to repeat all these calculations for the remaining cases, and hence we leave it as a modest exercise by the interested readers. The result of such symmetry hunting is summarized in Table~\ref{tab:helioP-symmetry}. For possible convenience of the readers, the relationship between the solutions given in Table~\ref{tab:helioP-symmetry} and the parameters of the SF equation is tabulated in Table~\ref{tab:SF-solutions}.

\begin{table}[h!]
\vglue -0.2cm
\begin{center}
\caption{ Summary of the reparametrization symmetry in the helio-perturbation theory~\cite{Minakata:2015gra}. The symmetry denoted as e.g., ``Symmetry X'' in this Table is called as ``Symmetry X-helioP'' in the text, where X = IA, IB, IIA, IIB, IIIA, IIIB, IVA, and IVB. Each type is paired by no ``f'' and with ``f'' which show without or with sign flipping of $s_{12}$, respectively.
}
\label{tab:helioP-symmetry}
\vglue 0.2cm
\begin{tabular}{c|c|c}
\hline 
Symmetry & 
Vacuum parameter transformations & 
Matter parameter transformations
\\
\hline 
\hline 
Symmetry IA & 
none & 
$\lambda_{-} \leftrightarrow \lambda_{+}$, 
$c_{\phi} \rightarrow \mp s_{\phi}$, 
$s_{\phi} \rightarrow \pm c_{\phi}$. \\
 & & 
$c_{(\phi - \theta_{13})} \rightarrow \mp s_{(\phi - \theta_{13})}$, 
$s_{(\phi - \theta_{13})} \rightarrow \pm c_{(\phi - \theta_{13})}$ \\
\hline 
Symmetry IAf & 
$\theta_{13} \rightarrow \theta_{13} \pm \pi$, $\theta_{12} \rightarrow - \theta_{12}$ & 
$\lambda_{-} \leftrightarrow \lambda_{+}$, 
$c_{\phi} \rightarrow \mp s_{\phi}$, 
$s_{\phi} \rightarrow \pm c_{\phi}$. \\
 & & 
$c_{(\phi - \theta_{13})} \rightarrow \pm s_{(\phi - \theta_{13})}$, 
$s_{(\phi - \theta_{13})} \rightarrow \mp c_{(\phi - \theta_{13})}$ \\
\hline 
Symmetry IB & 
$\theta_{13} \rightarrow - \theta_{13}$,  
$\delta \rightarrow \delta + \pi$. & 
$\lambda_{-} \leftrightarrow \lambda_{+}$, 
$c_{\psi} \rightarrow \pm s_{\psi}$, 
$s_{\psi} \rightarrow \pm c_{\psi}$. \\
 & & 
$c_{(\phi - \theta_{13})} \rightarrow \pm s_{(\phi - \theta_{13})}$, 
$s_{(\phi - \theta_{13})} \rightarrow \pm c_{(\phi - \theta_{13})}$ \\
\hline 
Symmetry IBf & 
$\theta_{13} \rightarrow - \theta_{13} \pm \pi $, 
$\theta_{12} \rightarrow - \theta_{12}$, & 
$\lambda_{-} \leftrightarrow \lambda_{+}$, 
$c_{\psi} \rightarrow \pm s_{\psi}$, 
$s_{\psi} \rightarrow \pm c_{\psi}$. \\
 & $\delta \rightarrow \delta + \pi$ & 
$c_{(\phi - \theta_{13})} \rightarrow \mp s_{(\phi - \theta_{13})}$, 
$s_{(\phi - \theta_{13})} \rightarrow \mp c_{(\phi - \theta_{13})}$ \\
\hline
Symmetry IIA & 
$\theta_{23} \rightarrow - \theta_{23}$, 
$\theta_{13} \rightarrow - \theta_{13}$ & 
$\lambda_{-} \leftrightarrow \lambda_{+}$, 
$c_{\phi} \rightarrow \pm s_{\phi}$, 
$s_{\phi} \rightarrow \pm c_{\phi}$ \\ 
 & & 
$c_{(\phi - \theta_{13})} \rightarrow \pm s_{(\phi - \theta_{13})}$, 
$s_{(\phi - \theta_{13})} \rightarrow \pm c_{(\phi - \theta_{13})}$ \\
\hline 
Symmetry IIAf & 
$\theta_{23} \rightarrow - \theta_{23}$, 
$\theta_{13} \rightarrow - \theta_{13} \pm \pi$ 
 & 
$\lambda_{-} \leftrightarrow \lambda_{+}$, 
$c_{\phi} \rightarrow \pm s_{\phi}$, 
$s_{\phi} \rightarrow \pm c_{\phi}$ \\ 
 & $\theta_{12} \rightarrow - \theta_{12}$  & 
$c_{(\phi - \theta_{13})} \rightarrow \mp s_{(\phi - \theta_{13})}$, 
$s_{(\phi - \theta_{13})} \rightarrow \mp c_{(\phi - \theta_{13})}$ \\
\hline 
Symmetry IIB & 
$\theta_{23} \rightarrow - \theta_{23}$, 
$\delta \rightarrow \delta + \pi$ & 
$\lambda_{-} \leftrightarrow \lambda_{+}$, 
$c_{\phi} \rightarrow \mp s_{\phi}$, 
$s_{\phi} \rightarrow \pm c_{\phi}$  \\ 
 & & 
$c_{(\phi - \theta_{13})} \rightarrow \mp s_{(\phi - \theta_{13})}$, 
$s_{(\phi - \theta_{13})} \rightarrow \pm c_{(\phi - \theta_{13})}$ \\
\hline 
Symmetry IIBf & 
$\theta_{23} \rightarrow - \theta_{23}$, 
$\theta_{13} \rightarrow \theta_{13} \pm \pi$, & 
$\lambda_{-} \leftrightarrow \lambda_{+}$, 
$c_{\phi} \rightarrow \mp s_{\phi}$, 
$s_{\phi} \rightarrow \pm c_{\phi}$  \\ 
 & $\theta_{12} \rightarrow - \theta_{12}$, $\delta \rightarrow \delta + \pi$ & 
$c_{(\phi - \theta_{13})} \rightarrow \pm s_{(\phi - \theta_{13})}$, 
$s_{(\phi - \theta_{13})} \rightarrow \mp c_{(\phi - \theta_{13})}$. \\
\hline 
Symmetry IIIA & 
$\theta_{13} \rightarrow - \theta_{13} \pm \pi$, & 
$\lambda_{-} \leftrightarrow \lambda_{+}$, 
$c_{\phi} \rightarrow \pm s_{\phi}$, 
$s_{\phi} \rightarrow \pm c_{\phi}$ \\ 
 & & 
$c_{(\phi - \theta_{13})} \rightarrow \mp s_{(\phi - \theta_{13})}$, 
$s_{(\phi - \theta_{13})} \rightarrow \mp c_{(\phi - \theta_{13})}$ \\
\hline 
Symmetry IIIAf & 
$\theta_{13} \rightarrow - \theta_{13}$, 
$\theta_{12} \rightarrow - \theta_{12}$. & 
$\lambda_{-} \leftrightarrow \lambda_{+}$, 
$c_{\phi} \rightarrow \pm s_{\phi}$, 
$s_{\phi} \rightarrow \pm c_{\phi}$ \\ 
 & & 
$c_{(\phi - \theta_{13})} \rightarrow \pm s_{(\phi - \theta_{13})}$, 
$s_{(\phi - \theta_{13})} \rightarrow \pm c_{(\phi - \theta_{13})}$ \\
\hline 
Symmetry IIIB & 
$\theta_{13} \rightarrow \theta_{13} \pm \pi$, 
$\delta \rightarrow \delta + \pi$. & 
$\lambda_{-} \leftrightarrow \lambda_{+}$, 
$c_{\phi} \rightarrow \mp s_{\phi}$, 
$s_{\phi} \rightarrow \pm c_{\phi}$ \\ 
 & & 
$c_{(\phi - \theta_{13})} \rightarrow \pm s_{(\phi - \theta_{13})}$, 
$s_{(\phi - \theta_{13})} \rightarrow \mp c_{(\phi - \theta_{13})}$ \\
\hline 
Symmetry IIIBf & 
$\theta_{12} \rightarrow - \theta_{12}$, 
$\delta \rightarrow \delta + \pi$. & 
$\lambda_{-} \leftrightarrow \lambda_{+}$, 
$c_{\phi} \rightarrow \mp s_{\phi}$, 
$s_{\phi} \rightarrow \pm c_{\phi}$ \\ 
 & & 
$c_{(\phi - \theta_{13})} \rightarrow \mp s_{(\phi - \theta_{13})}$, 
$s_{(\phi - \theta_{13})} \rightarrow \pm c_{(\phi - \theta_{13})}$ \\
\hline 
Symmetry IVA & 
$\theta_{23} \rightarrow - \theta_{23}$, 
$\theta_{13} \rightarrow \theta_{13} \pm \pi$ & 
$\lambda_{-} \leftrightarrow \lambda_{+}$, 
$c_{\phi} \rightarrow \mp s_{\phi}$, 
$s_{\phi} \rightarrow \pm c_{\phi}$ \\ 
 & & 
$c_{(\phi - \theta_{13})} \rightarrow \pm s_{(\phi - \theta_{13})}$, 
$s_{(\phi - \theta_{13})} \rightarrow \mp c_{(\phi - \theta_{13})}$ \\
\hline 
Symmetry IVAf & 
$\theta_{23} \rightarrow - \theta_{23}$, 
$\theta_{12} \rightarrow - \theta_{12}$ & 
$\lambda_{-} \leftrightarrow \lambda_{+}$, 
$c_{\phi} \rightarrow \mp s_{\phi}$, 
$s_{\phi} \rightarrow \pm c_{\phi}$ \\ 
 & & 
$c_{(\phi - \theta_{13})} \rightarrow \mp s_{(\phi - \theta_{13})}$, 
$s_{(\phi - \theta_{13})} \rightarrow \pm c_{(\phi - \theta_{13})}$ \\
\hline 
Symmetry IVB & 
$\theta_{23} \rightarrow - \theta_{23}$, 
$\theta_{13} \rightarrow - \theta_{13} \pm \pi$, & 
$\lambda_{-} \leftrightarrow \lambda_{+}$, 
$c_{\phi} \rightarrow \pm s_{\phi}$, 
$s_{\phi} \rightarrow \pm c_{\phi}$ \\ 
 &
$\delta \rightarrow \delta + \pi$. 
 &
$c_{(\phi - \theta_{13})} \rightarrow \mp s_{(\phi - \theta_{13})}$, 
$s_{(\phi - \theta_{13})} \rightarrow \mp c_{(\phi - \theta_{13})}$ \\
\hline 
Symmetry IVBf & 
$\theta_{23} \rightarrow - \theta_{23}$, 
$\theta_{13} \rightarrow - \theta_{13}$, & 
$\lambda_{-} \leftrightarrow \lambda_{+}$, 
$c_{\phi} \rightarrow \pm s_{\phi}$, 
$s_{\phi} \rightarrow \pm c_{\phi}$ \\ 
 &
$\theta_{12} \rightarrow - \theta_{12}$, $\delta \rightarrow \delta + \pi$. 
 &
$c_{(\phi - \theta_{13})} \rightarrow \pm s_{(\phi - \theta_{13})}$, 
$s_{(\phi - \theta_{13})} \rightarrow \pm c_{(\phi - \theta_{13})}$ \\
\hline 
\end{tabular}
\end{center}
\vglue -0.4cm 
\end{table}

\begin{table}[h!]
\vglue -0.2cm
\begin{center}
\caption{ 
The relationship between the solutions given in Table~\ref{tab:helioP-symmetry} and the parameters of the SF equation. The labels ``upper'' and ``lower'' imply the upper and lower sign in the corresponding columns in Table~\ref{tab:helioP-symmetry}. 
}
\label{tab:SF-solutions}
\vglue 0.2cm
\begin{tabular}{c|c|c}
\hline 
Symmetry & 
$\tau, \sigma, \xi$ & 
$\alpha, \beta$
\\
\hline 
\hline 
Symmetry IA, IAf & 
$\tau = 0$, $ \sigma = 0$, $\xi = 0$ & 
$\alpha = \beta = 0$ (upper) \\ 
& &
$\alpha = \pi, \beta = - \pi$ (lower)  \\
\hline 
Symmetry IB, IBf & 
$\tau = \sigma = 0$, $\xi = \pi$ & 
$\alpha = \pi, \beta = - \pi$ (upper) \\
& & $\alpha = \beta = 0$ (lower) \\
\hline 
Symmetry IIA, IIAf & 
$\tau = 0, \sigma = - \pi$, $\xi = 0$ & 
$\alpha = \pi, \beta = 0$ (upper) \\
& & $\alpha = 0, \beta = \pi$ (lower) \\
\hline 
Symmetry IIB, IIBf & 
$\tau = 0, \sigma = - \pi$, $\xi = \pi$ & 
$\alpha = 0, \beta = \pi$ (upper) \\ 
& & $\alpha = \pi, \beta = 0$ (lower) \\
\hline 
Symmetry IIIA, IIIAf & 
$\tau = \pi$, $\sigma = 0$, $\xi = 0$ & 
$\alpha = 0, \beta = \pi$ (upper) \\ 
& & $\alpha = \pi, \beta = 0$ (lower) \\
\hline 
Symmetry IIIB, IIIBf & 
$\tau = \pi, \sigma = 0$, $\xi = \pi$ & 
$\alpha = \pi, \beta = 0$ (upper)  \\ 
 & & 
$\alpha = 0, \beta = \pi$ (lower)   \\
\hline 
Symmetry IVA, IVAf & 
$\tau = \sigma = \pi$, $\xi = 0$ & 
$\alpha = \pi, \beta = - \pi$ (upper) \\ 
& &
$\alpha = \beta = 0$ (lower)  \\
\hline 
Symmetry IVB, IVBf & 
$\tau = \pi$, $\sigma = \pi$, $\xi = \pi$ & 
$\alpha = \beta = 0$ (upper) \\ 
& &
$\alpha = \pi, \beta = - \pi$ (lower)  \\
\hline 
\end{tabular}
\end{center}
\vglue -0.4cm 
\end{table}

The structure and variety of symmetry in the helio-perturbation theory is very similar to that of the DMP case~\cite{Minakata:2021dqh}. We have the four types, I, II, III, and IV, doubled with Types A (no $\delta$) and B (with $\delta$), in which only Symmetry IA and IB are free free from the flavor basis rephasing. However, a clear difference exists. That is a new pairing of symmetries with or without $\theta_{12}$ sign flip, which entailed the doubled, sixteen symmetries in the helio-perturbation theory as opposed to the eight in DMP. Another difference is, of course, here we discuss the 1-3 state exchange symmetry, but in DMP the 1-2 state exchange one, whose two levels nearly cross in very different kinematical regions. 

Interestingly, Table~\ref{tab:SF-solutions} for the solutions to the SF equation is {\em identical} with Table~2 for the DMP symmetries~\cite{Minakata:2021dqh} apart from the presence or absence of the $s_{12}$ flip solutions. It suggests that the structure I-IV doubled with Types A and B represents the universal feature of state-relabeling symmetries in neutrino oscillation in matter.

One may ask: Why so many, doubled number of symmetries compared to DMP? The DMP perturbation theory is a global framework for all the terrestrial experiments which include the both atmospheric-scale and solar-scale resonances~\cite{Denton:2016wmg}, as emphasized in ref.~\cite{Minakata:2020oxb}. For a pictorial view, see e.g., Fig.~1 in ref.~\cite{Minakata:2019gyw}. Even if we discuss physics in region of the atmospheric-scale enhanced oscillation, the variables which describe the solar-scale oscillation are not the spectators. On the other hand, in the helio-perturbation theory, the region of solar-scale oscillation is outside the region of validity of the theory. Therefore, the solar variables are completely spectators. That is probably why the symmetries in the helio-perturbation theory is doubled due to the $s_{12}$ sign flip and non-flip dualism. 

The characteristic feature which one can observe through the process of symmetry finding and is worth to comment here again is the tight relationship between the vacuum- and the matter-variables transformations. It must be obvious from the treatment of sections~\ref{sec:IA-helioP} and \ref{sec:IB-helioP} that the transformations of $\phi$, $\theta_{13}$, and $\theta_{12}$ are all related to each others to organize themselves into the symmetries IA, IAf, IB, and IBf in the helio-perturbation theory. The similar conspiracy of the vacuum and matter variables appeared in our treatment of the SF equation in DMP in which, for example, only one of the $s_{12}$ sign flip, or non-flip solution is allowed depending upon the transformations of the other vacuum and matter parameters. See ref.~\cite{Minakata:2021dqh}. 

\section{Hamiltonian view of the 1-3 exchange symmetry in matter} 
\label{sec:hamiltonian-view} 

We have already introduced in section~\ref{sec:3nu-SM} the Hamiltonian view of the symmetry. There are two ways of expressing the flavor basis Hamiltonian, the originally defined form $H_{ \text{LHS} } = H_{ \text{vac} } + H_{ \text{matt} }$, and its diagonalized form $H_{ \text{RHS} }$. Of course, they must be equal to each other. 

In the helio-perturbation theory $H_{ \text{RHS} }$ is not exactly diagonalized, but decomposed into the unperturbed (zeroth-order) and perturbed (first-order) Hamiltonian. In the PDG convention $2E H _{ \text{RHS} }$ is given by 
\begin{eqnarray} 
&& 
2E H _{ \text{RHS} } 
= U_{23} (\theta_{23}) U_{13} (\phi, \delta) 
\nonumber \\
&\times&
\left\{ 
\left[
\begin{array}{ccc}
\lambda_{-} & 0 & 0 \\
0 & \lambda_{0} & 0 \\
0 & 0 & \lambda_{+}
\end{array}
\right] 
+ 
\epsilon c_{12} s_{12} \Delta m^2_{\rm ren} 
\left[
\begin{array}{ccc}
0 & c_{ (\phi - \theta_{13}) } & 0 \\
c_{ (\phi - \theta_{13}) } & 0 & s_{ (\phi - \theta_{13}) } e^{- i \delta} \\
0  & s_{ (\phi - \theta_{13}) } e^{ i \delta} & 0 
\end{array}
\right] 
\right\} 
U_{13}^{\dagger} (\phi, \delta) U_{23}^{\dagger} (\theta_{23}).
\nonumber \\
\label{H-RHS-helioP}
\end{eqnarray}
Of course $H_{ \text{LHS} }$ is the same as in eq.~\eqref{H-LHS}.

In this section we show that all the symmetries derived in section~\ref{sec:13-symmetry-result} are the Hamiltonian symmetries. That is, the transformations belonging to each symmetry tabulated in Table~\ref{tab:helioP-symmetry} leave $H_{ \text{LHS} }$ and $H_{ \text{RHS} }$ in eq.~\eqref{H-RHS-helioP} invariant up to a common rephasing matrix. Since there are so many symmetries, sixteen of them, which does not quite fit to the explicit treatment for demonstrating the invariance here, we pick only one of them to show the point. But, the interested readers can easily work out the invariance for the rest of the symmetries. 

\subsection{Symmetry IVB-helioP}
\label{sec:IVB} 

We pick up Symmetry IVB-helioP (see Table~\ref{tab:helioP-symmetry}, the second from the bottom) for an example for explicit demonstration of invariance of the Hamiltonian. We first discuss invariance of $H_{ \text{LHS} }$. 
The vacuum parameter transformations in Symmetry IVB-helioP include: 
$\theta_{23} \rightarrow - \theta_{23}$, $\theta_{13} \rightarrow - \theta_{13} \pm \pi$, and $\delta \rightarrow \delta + \pi$. $\theta_{13} \rightarrow - \theta_{13} \pm \pi$ implies $c_{13} \rightarrow - c_{13}$ and $s_{13} \rightarrow s_{13}$. Then, under these transformations the $U=U_{23} (\theta_{23}) U_{13} (\theta_{13}, \delta) U_{12} (\theta_{12})$ matrix transforms under Symmetry IVB-helioP as\footnote{
The transformation property of the $U$ matrix with the asymmetric rephasing matrix, from left only, is new. It has never showed up in the case of the DMP symmetries discussed in ref.~\cite{Minakata:2021dqh}. The similar asymmetric rephasing factors on $U$ also appear in the treatment of Symmetry IIAf, IIBf, IIIA, IIIB, IVA. } 
\begin{eqnarray} 
&& 
U \rightarrow 
\text{Rep(IV)} U,  
\hspace{10mm}
\text{Rep(IV)} 
\equiv 
\left[
\begin{array}{ccc}
- 1 & 0 & 0 \\
0 & 1 & 0 \\
0 & 0 & -1
\end{array}
\right]. 
\label{U-transform}
\end{eqnarray}
The rephasing matrix - Rep(IV) (``-'' means minus) can do the job as well. Since $H_{ \text{vac} } = U \text{diag} (m^2_{1}, m^2_{2}, m^2_{3}) U^{\dagger}$, $H_{ \text{vac} }$ and hence $H_{ \text{LHS} }$ is invariant under Symmetry IVB-helioP transformations up to the Rep(IV) rephasing matrix, which is operated both from the left and the right.

Now we discuss invariance of $H_{ \text{RHS} }$. By $c_{\phi} \rightarrow \pm s_{\phi}$, $s_{\phi} \rightarrow \pm c_{\phi}$ and $\delta \rightarrow \delta + \pi$, $U_{13} (\phi, \delta)$ transforms as  
\begin{eqnarray} 
&& 
U_{13} (\phi, \delta) 
= 
\left[
\begin{array}{ccc}
c_{\phi}  & 0 &  s_{\phi} e^{- i \delta} \\
0 & 1 & 0 \\
- s_{\phi} e^{ i \delta} & 0 & c_{\phi}  \\
\end{array}
\right] 
\rightarrow 
U_{13} (\phi^{\prime}, \delta^{\prime}) 
=
\left[
\begin{array}{ccc}
\pm s_{\phi} & 0 &  \mp c_{\phi} e^{- i \delta} \\
0 & 1 & 0 \\
\pm c_{\phi} e^{ i \delta} & 0 & \pm s_{\phi} \\
\end{array}
\right] 
\nonumber 
\end{eqnarray}
where $\phi^{\prime} = - \phi \pm \frac{\pi}{2}$, and $\delta^{\prime} = \delta + \pi$. Then, 
\begin{eqnarray} 
&& 
\hspace{-6mm}
U_{13} (\phi, \delta) 
\left[
\begin{array}{ccc}
\lambda_{-} & 0 & 0 \\
0 & \lambda_{0} & 0 \\
0 & 0 & \lambda_{+}
\end{array}
\right]
U_{13}^{\dagger} (\phi, \delta) 
\rightarrow
U_{13} (\phi^{\prime}, \delta^{\prime}) 
\left[
\begin{array}{ccc}
\lambda_{+} & 0 & 0 \\
0 & \lambda_{0} & 0 \\
0 & 0 & \lambda_{-}
\end{array}
\right]
U_{13}^{\dagger} (\phi^{\prime}, \delta^{\prime}) 
\nonumber \\
&& 
\hspace{-10mm}
=
\left[
\begin{array}{ccc}
s_{\phi}^2 \lambda_{+} + c_{\phi}^2 \lambda_{-} & 0 & 
c_{\phi} s_{\phi} e^{- i \delta} ( \lambda_{+} - \lambda_{-} ) \\
0 & \lambda_{0} & 0 \\
c_{\phi} s_{\phi} e^{ i \delta} ( \lambda_{+} - \lambda_{-} ) & 0 & 
c_{\phi}^2 \lambda_{+} + s_{\phi}^2 \lambda_{-}  \\
\end{array}
\right] 
=
U_{13} (\phi, \delta) 
\left[
\begin{array}{ccc}
\lambda_{-} & 0 & 0 \\
0 & \lambda_{0} & 0 \\
0 & 0 & \lambda_{+}
\end{array}
\right]
U_{13}^{\dagger} (\phi, \delta).
\label{H0-transf}
\end{eqnarray}
under the transformations of Symmetry IVB-helioP. Therefore, $H_{0}$ is invariant by itself, without the Rep(IV) rephasing matrix. It may look like a trouble but it is not. One can easily show that $H_{0}$, the RHS of eq.~\eqref{H0-transf} is invariant under multiplication of the Rep(IV) rephasing matrix from left and right. 

The matrix part in the perturbed Hamiltonian $H_{1}$ transforms under Symmetry IVB-helioP transformations 
\begin{eqnarray} 
&& 
U_{13} (\phi, \delta) 
\left[
\begin{array}{ccc}
0 & c_{ (\phi - \theta_{13}) } & 0 \\
c_{ (\phi - \theta_{13}) } & 0 & s_{ (\phi - \theta_{13}) } e^{- i \delta} \\
0  & s_{ (\phi - \theta_{13}) } e^{ i \delta} & 0 
\end{array}
\right] 
U_{13}^{\dagger} (\phi, \delta) 
\nonumber \\
&\rightarrow& 
U_{13} (\phi^{\prime}, \delta^{\prime}) 
\left[
\begin{array}{ccc}
0 & \mp s_{ (\phi - \theta_{13}) } & 0 \\
\mp s_{ (\phi - \theta_{13}) } & 0 & \pm c_{ (\phi - \theta_{13}) } e^{- i \delta} \\
0  & \pm c_{ (\phi - \theta_{13}) } e^{ i \delta} & 0 
\end{array}
\right] 
U_{13}^{\dagger} (\phi^{\prime}, \delta^{\prime}) 
\nonumber \\
&=& 
- U_{13} (\phi, \delta) 
\left[
\begin{array}{ccc}
0 & c_{ (\phi - \theta_{13}) } & 0 \\
c_{ (\phi - \theta_{13}) } & 0 & s_{ (\phi - \theta_{13}) } e^{- i \delta} \\
0  & s_{ (\phi - \theta_{13}) } e^{ i \delta} & 0 
\end{array}
\right] 
U_{13}^{\dagger} (\phi, \delta).
\label{H1-transf}
\end{eqnarray}
While the minus sign looks troublesome but actually not. The structure of non-vanishing lozenge position elements is maintained in the all three terms in eq.~\eqref{H1-transf}. Hence, the minus sign can be cancelled when the Rep(IV) rephasing matrices are multiplied from left and right. Notice that there is no $s_{12}$ sign flip in Symmetry IVB-helioP. Therefore, $H_{1}$ is invariant under Symmetry IVB-helioP transformations up to the Rep(IV) rephasing. 

The last step to prove the invariance of $H_{1}$ is to show that the Rep(IV) rephasing matrices successfully exit from the above $U_{13}^{\dagger} (\phi, \delta) \mathcal{O} U_{13}^{\dagger} (\phi, \delta)$ part to left and right of $H_{1}$, see $H _{ \text{RHS} }$ in eq.~\eqref{H-RHS-helioP}. Notice that the transformation $\theta_{23} \rightarrow - \theta_{23}$ transformation is involved in Symmetry IVB-helioP. It is easy to see how it can be done: 
\begin{eqnarray} 
&& 
U_{23} ( - \theta_{23}) 
\text{Rep(IV)} 
U_{13}^{\dagger} (\phi, \delta) \mathcal{O} U_{13}^{\dagger} (\phi, \delta) 
\text{Rep(IV)} 
U_{23}^{\dagger} ( - \theta_{23}) 
\nonumber \\
&=& 
\text{Rep(IV)} 
U_{23} ( \theta_{23}) 
U_{13}^{\dagger} (\phi, \delta) \mathcal{O} U_{13}^{\dagger} (\phi, \delta) 
U_{23}^{\dagger} ( \theta_{23}) 
\text{Rep(IV)}, 
\label{pass-through}
\end{eqnarray}
where we have used the property $\text{Rep(IV)} U_{23} ( - \theta_{23}) \text{Rep(IV)} = U_{23} ( \theta_{23})$. That is, passing through the rephasing matrix remedies the flipped sign of $\sin \theta_{23}$. 

To summarize, the both $H_{0}$ and $H_{1}$, hence $H_{ \text{RHS} }$ is invariant under Symmetry IVB-helioP transformations with the Rep(IV) rephasing factor which is identical with the one for $H_{ \text{LHS} }$. It implies that Symmetry IVB-helioP is the Hamiltonian symmetry. 

\subsection{All the rest of symmetries in the helio-perturbation theory}
\label{sec:all-rest} 

One can repeat the similar analysis for all the symmetries tabulated in Table~\ref{tab:helioP-symmetry} to prove that they are all Hamiltonian symmetries. First, one has to go through the analysis of how $H_{ \text{LHS} }$ transforms under Symmetry X, where X=I, II, III, IV, and identifies the rephasing matrix Rep(X). As in the case of DMP, Rep(X) depends only on the type X=I, II, III, IV, and independent of Types A or B, and ``f'' type or non ``f'' type symmetries. Rep(IV) is already given in eq.~\eqref{U-transform}.

Next, one should examine the transformation properties of $H_{ \text{RHS} }$ under Symmetry X. It involves the both $H_{0}$ and $H_{1}$, the first and second terms in eq.~\eqref{H-RHS-helioP}. One can show that $H_{0}$ and $U_{13}^{\dagger} (\phi, \delta) \mathcal{O} U_{13}^{\dagger} (\phi, \delta)$ part of $H_{1}$ are invariant under Symmetry X up to the same rephasing matrix obtained from $H_{ \text{LHS} }$ transformation analysis. 
The required rephasing matrices depend on symmetry types (see Table~\ref{tab:helioP-symmetry}), 
\begin{eqnarray} 
&& 
\text{Rep(II)} 
\equiv 
\left[
\begin{array}{ccc}
1 & 0 & 0 \\
0 & 1 & 0 \\
0 & 0 & -1
\end{array}
\right] ~~\text{for~ IIA, IIAf, IIB, IIBf}, ~~
\nonumber \\
&& 
\text{Rep(III)} 
\equiv 
\left[
\begin{array}{ccc}
-1 & 0 & 0 \\
0 & 1 & 0 \\
0 & 0 & 1
\end{array}
\right] ~~\text{for~ IIIA, IIIAf, IIIB, IIIBf}, 
\nonumber \\
&& 
\text{Rep(IV)} 
\equiv 
\left[
\begin{array}{ccc}
- 1 & 0 & 0 \\
0 & 1 & 0 \\
0 & 0 & - 1
\end{array}
\right] ~~\text{for~ IVA, IVAf, IVB, IVBf},
\label{rephasing}
\end{eqnarray}
and no rephasing matrix is needed for IA, IAf, IB, IBf. 
In the ``f'' type symmetries in which $s_{12}$ sign flip is involved, the ``troublesome minus sign'' remains in the $H_{1}$ transformation similar to eq.~\eqref{H1-transf} even after the same rephasing factor as in $H_{ \text{LHS} }$ is introduced. But, this minus sign is cancelled by another minus sign which comes from $s_{12}$ sign flip.

The remaining issue is, then, how passing-through rephasing matrix affects $U_{23} ( \pm \theta_{23})$, the similar problem discussed in eq.~\eqref{pass-through}. We note the property of the rephasing matrix 
\begin{eqnarray} 
&&
\text{Rep(X)} U_{23} ( - \theta_{23}) \text{Rep(X)} = U_{23} ( \theta_{23}), 
\hspace{10mm} 
\text{X=II, IV,} 
\nonumber \\
&&
\text{Rep(III)} U_{23} ( \theta_{23}) \text{Rep(III)} = U_{23} ( \theta_{23}).
\label{passing-Rep}
\end{eqnarray}
For Symmetry X=II and IV, the rephasing matrix has to pass through $U_{23} ( - \theta_{23})$ because $\theta_{23} \rightarrow - \theta_{23}$ transformation is involved,  see Table~\ref{tab:helioP-symmetry}. For Symmetry X=III, no $\theta_{23}$ transformation is involved, so that Rep(III) passes through $U_{23} ( \theta_{23})$. The formulas in \eqref{passing-Rep} imply that the rephasing matrix exits from the $U_{13}^{\dagger} (\phi, \delta) \mathcal{O} U_{13}^{\dagger} (\phi, \delta)$ part of $H_{1}$  with remedying the minus sign of $U_{23} ( - \theta_{23})$ for Symmetry X=II and IV, and just pass through $U_{23} ( \theta_{23})$ without affecting it for Symmetry III. Therefore, the both $H_{ \text{LHS} }$ and $H_{ \text{RHS} }$ are invariant under all Symmetry X (X=I, II, III, IV), apart from the rephasing matrix given in eq.~\eqref{rephasing}.

Thus, all the sixteen symmetries tabulated in Table~\ref{tab:helioP-symmetry} are the Hamiltonian symmetries. What is good for a Hamiltonian symmetry is that 
(1) the symmetry holds to all orders of perturbation theory, and 
(2) it is valid for varying density matter profile, the properties pointed out in ref.~\cite{Minakata:2021dqh} for the DMP symmetries. 

Finally, we should make a remark on possible ``Naumov test'' for the symmetries we have obtained in this section. Unfortunately, there is no Naumov identity in the helio-perturbation theory. The reason for this property is explained in Appendix~\ref{sec:Naumov-helioP}. 

\section{Concluding remarks}
\label{sec:conclusion}

In this paper, we have reported a new progress in uncovering the state relabeling symmetry by using the Symmetry Finder (SF) method~\cite{Minakata:2021dqh} in neutrino oscillations in matter. That is, the sixteen 1-3 state exchange symmetries came out in the helio-perturbation theory, which describes the region of atmospheric-scale enhanced oscillation. It is the second application of the SF method after the similar treatment of the 1-2 state exchange symmetry in DMP, which produced the eight symmetries~\cite{Minakata:2021dqh}. These results testify that the SF method is flexible, easy to use, and sufficiently powerful to identify symmetries in neutrino oscillation in matter. 

The structure and variety of symmetry in the helio-perturbation theory is very similar to that of DMP, apart from the doubling due to $s_{12}$ sign flip and non-flip dualism. The DMP symmetries have the four types, I, II, III, and IV, doubled with Types A and B, where they are distinguished by without (A) and with (B) $\delta$ transformations, the structure common to the helio-perturbation theory. They constitute the eight symmetries, which is further doubled by $s_{12}$ sign flip or non-flip, supplying another eight in the helio-perturbation theory, the only different aspect of symmetry between the two theories. 

Why the symmetry is doubled in the helio-perturbation theory? The DMP perturbation theory is constructed in such a way that the whole region covered by the terrestrial neutrino experiments is inside the region of validity of the theory~\cite{Denton:2016wmg,Minakata:2020oxb}. Whereas in the helio-perturbation theory, it is outside the region of validity of the theory, and hence the solar variables are completely spectators in the helio-perturbation theory, unlike in DMP. That is probably why the symmetries in the helio-perturbation theory is doubled due to the availability of $s_{12}$ sign flip and non-flip options. 

In section~\ref{sec:overview}, we have investigated the two-flavor model to make clear the relationship between the vacuum symmetry (of the Hamiltonian $H_{ \text{LHS} }$) and matter symmetry (of the Hamiltonian $H_{ \text{RHS} }$). We have observed a rather intricate structure. While Symmetry IA-vacuum is the symmetry of the system in any finite order in perturbation theory based on $H_{ \text{LHS} }$, if the perturbative series is summed to all orders, the system approaches to the diagonalized one in a non-perturbative fashion, which has Symmetry IA-ZS. Therefore, it appears that the vacuum symmetry fuses into the matter symmetry through the process of all-order summation. We would like to note that the question of the relationship between the vacuum symmetry and matter symmetry is even more nontrivial one with the three generation neutrinos. The Toshev identity indicates that the vacuum-symmetry and matter-symmetry transformations cannot be independent in the $\theta_{23}$-$\delta$ and $\widetilde{\theta_{23}}$-$\widetilde{\delta}$ sub-secter. 

A more general approach to symmetries including the state exchange and non-exchange types is taken by the authors of ref.~\cite{Denton:2021vtf}. In our SF method, since the vacuum-variable transformations are tightly linked to the matter-variable ones, the direct product structure of symmetries of $H_{ \text{LHS} }$ and $H_{ \text{RHS} }$ is unlikely to come out. But, we feel it better to wait their final report to discuss the relationship between the two approaches, because the characteristically different new symmetries are expected to exist in their framework~\cite{Denton:2021vtf}. 

Finally, we would like to repeat our interest in the possibility that the neutrino theory under the matter potential can be viewed as a mean-field treatment of self-interacting neutrino system apart from the difference between electron and neutrino density potentials~\cite{Minakata:2021dqh}. It would be an interesting problem to examine what is the consequence of the state exchange symmetries written by the dynamical variables.

\appendix

\section{Naumov identity in the helio-perturbation theory}
\label{sec:Naumov-helioP}

The Naumov identity has a problem in the helio-perturbation theory, so that let us understand it first. In this appendix we use the ATM convention of the $U$ matrix not to worry about the change into the PDG convention. Naively, the Naumov identity may be written in the helio-perturbation theory as 
\begin{eqnarray} 
&& 
c_{23} s_{23} c^2_{13} s_{13} c_{12} s_{12} \sin \delta ~ 
( m^2_{2} - m^2_{1} ) ( m^2_{3} - m^2_{2} ) ( m^2_{3} - m^2_{1} ) 
\nonumber \\
&\approx& 
c_{23} s_{23} c^2_{\phi} s_{\phi} c_{12} s_{12} \sin \delta ~
( \lambda_{2} - \lambda_{1} ) ( \lambda_{3} - \lambda_{2} )( \lambda_{3} - \lambda_{1} ), ~~
\label{Naumov-helioP}
\end{eqnarray}
where the leading order expression is to be inserted in the RHS of eq.~\eqref{Naumov-helioP}. We assume the normal mass ordering and neutrino channel for which $\lambda_{1}$, $\lambda_{2}$, $\lambda_{3}$ correspond to $\lambda_{0}$, $\lambda_{-}$, $\lambda_{+}$, respectively. 

By using the leading order expressions of the eigenvalues, $c_{\phi}$ and $s_{\phi}$~\cite{Minakata:2015gra}, where $\phi$ is the $\theta_{13}$ in matter, one can write the RHS of eq.~\eqref{Naumov-helioP} as 
\begin{eqnarray} 
&& 
\sqrt{2} 
c_{23} s_{23} c^2_{13} s^2_{13} c_{12} s_{12} \sin \delta 
( \Delta m^2_{ \text{ren} } )^2 
\frac{ \left[ 
c^2_{13} a \Delta m^2_{ \text{ren} } 
- \epsilon \cos 2\theta_{12}  \Delta m^2_{ \text{ren} } 
\left( \Delta m^2_{ \text{ren} } + a \right) 
\right]  }{ \sqrt{ ( \lambda_{+} - \lambda_{-} ) \left[ ( \lambda_{+} - \lambda_{-} ) - \Delta m^2_{ \text{ren} } \cos 2\theta_{13} - a \right] } }.  
\nonumber \\
\label{RHS-helioP}
\end{eqnarray}
This result implies that the Naumov identity fails in the helio-perturbation theory, even as the approximate relation due to our limitation to the leading order expression in the RHS. The reason is that while the LHS of eq.~\eqref{Naumov-helioP} is of order $\Delta m^2_{21} (\Delta m^2_{31} )^2$, the RHS in eq.~\eqref{RHS-helioP} is of order $(\Delta m^2_{31} )^3$, no $\epsilon$ suppression. Of course, this is the common feature of all the versions of the helio-perturbation theory~\cite{Arafune:1996bt,Cervera:2000kp,Freund:2001pn,Akhmedov:2004ny,Minakata:2015gra}. 

The fact that the Naumov identity is not satisfied in the helio-perturbation theory should not be interpreted as a failure of the theory. The region of validity of the helio-perturbation theory is restricted to the region around the atmospheric resonance. In this region all the three eigenvalue differences are of order $\Delta m^2_{31}$, and the matter mixing angle $\phi$ is large to describe the ``resonance''. Therefore, there is no way to produce $\epsilon$ suppression in the RHS of eq.~\eqref{Naumov-helioP}. 

Of course, the Naumov identity holds in the exact ZS theory and in the DMP theory~\cite{Denton:2016wmg} at least approximately, whose region of validity spans a global region which include the solar and atmospheric resonance regions. In the DMP perturbation theory, the $\epsilon$ suppression in region of the atmospheric-scale enhanced oscillations is provided by the smallness of $\psi$, $c_{\psi} s_{\psi} \propto \epsilon$~\cite{Denton:2016wmg}. Therefore, it appears that the Naumov identity classifies the exact and approximate frameworks of neutrino oscillation into the two categories, with the global region of validity, the ZS~\cite{Zaglauer:1988gz} and DMP~\cite{Denton:2016wmg},\footnote{
It is likely that Agarwalla {\it et al.} (AKT) perturbation theory~\cite{Agarwalla:2013tza} is to be added to the list of the framework with global region of validity, as DMP and AKT are the two sister papers employing the Jacobi method. Despite clear interests, the SF method has not been applied to the AKT perturbation theory. 
}
and with the local region of validity, such as the helio-perturbative theory.

\end{document}